\documentclass[iop, twocolumn, tighten]{emulateapj}
\usepackage{amsmath}
\usepackage{color}
\definecolor{redgray}{rgb}{0.2,0.0,0.0}
\definecolor{usui}{rgb}{0.7,0.7,0.7} 

\newcommand{\NL}{\mathrm{null}}

\newcommand{\Nnull}{N_\mathrm{null}}
\newcommand{\Ndata}{N}
\newcommand{\Mmodel}{M}

\newcommand{\m}{{\boldsymbol m}}

\newcommand{\mest}{{\boldsymbol m}_{\mathrm{est}}}
\newcommand{\mestzt}{{\boldsymbol m}_{\mathrm{est}}^{\zeta,\Theta_S}}
\newcommand{\sing}{l}
\newcommand{\Lp}{\tilde{\Lambda}}

\newcommand{\A}{{\boldsymbol a}}
\newcommand{\Adata}{{\boldsymbol a}_\mathrm{data}}

\newcommand{\zetain}{\zeta_{\mathrm{input}}}
\newcommand{\ThetaSin}{\Theta_{S,\mathrm{input}}}
\newcommand{\zetaest}{\zeta_{\mathrm{est}}}
\newcommand{\ThetaSest}{\Theta_{S,\mathrm{est}}}
\newcommand{\SVI}{S_\mathrm{VI}}

\slugcomment{}
\shortauthors{Kawahara and Fujii}
\shorttitle{Global Mapping of Exoplanets}

\begin{document}
\title{Global Mapping of Earth-like Exoplanets from Scattered Light Curves}


\author{Hajime Kawahara\altaffilmark{1,2}
 and Yuka Fujii\altaffilmark{2}} 
\altaffiltext{1}{Department of Physics, Tokyo Metropolitan University,
  Hachioji, Tokyo 192-0397, Japan}
\altaffiltext{2}{Department of Physics, The University of Tokyo, 
Tokyo 113-0033, Japan}
\email{kawa\_h@tmu.ac.jp}

\begin{abstract}
Scattered lights from terrestrial exoplanets provide valuable information about the planetary surface. Applying the surface reconstruction method proposed by \cite{Fujii2010} to both diurnal and annual variations of the scattered light, we develop a reconstruction method of land distribution with both longitudinal and latitudinal resolutions. We find that one can recover a global map of an idealized Earth-like planet on the following assumptions: 1) cloudless, 2) a face-on circular orbit, 3) known surface types and their reflectance spectra 4) no atmospheric absorption, 5) known rotation rate 6) static map, and 7) no moon. Using the dependence of light curves on the planetary obliquity, we also show that the obliquity can be measured by adopting the $\chi^2$ minimization or the extended information criterion. We demonstrate a feasibility of our methodology by applying it to a multi-band photometry of a cloudless model Earth with future space missions such as the occulting ozone observatory (O3). We conclude that future space missions can estimate both the surface distribution and the obliquity at least for cloudless Earth-like planets within 5 pc. 
\end{abstract}

\section{Introduction}
 Recent progress in observational techniques has revealed various physical properties of exoplanets beyond orbital parameters and planetary mass. Detections of the atmospheric components have been reported for several systems using spectroscopy at the planetary transit and secondary eclipse \citep[e.g.][]{charbonneau2002, vidal2003, vidal2004, tinetti2007, swain2008, swain2009}. Interior compositions can be inferred from planetary mass and radius \citep[e.g.][]{2009A&A...506..287L, 2009Natur.462..891C}. Constructions of thermal maps of the planetary atmosphere have been proposed by \citep{2006ApJ...649.1020W, 2008ApJ...678L.129C}. A longitudinal thermal map of HD 189733b has been constructed by \cite{2007Natur.447..183K} based on the method proposed by \cite{2008ApJ...678L.129C}.
 
Nevertheless, an identification of planetary surface components still remains an ambitious challenge. One of the promising approaches is to use the scattered light of exoplanets through the direct imaging observation \citep[e.g.][]{2000ApJ...540..504S, 2001Natur.412..885F, 2005ApJ...627..520S}. \cite{2001Natur.412..885F} focused on the inhomogeneity of the Earth surface which causes diurnal variation of the scattered light. They computed the scattered light from a model Earth observed at a distance of 10 pc and showed that time variations of the scattered light in different photometric bands highly depend on the geological and biological features on the planetary surface such as ocean, land, and even vegetation. More detailed characterizations (including spectroscopy) of the scattered light of the Earth and its time variations are discussed both via Earth-shine observations \citep[e.g.][]{woolf2002, arnold2002,2006ApJ...651..544M} and simulations \citep{2006AsBio...6...34T, 2006AsBio...6..881T, 2006ApJ...651..544M}. These studies have suggested a future possibility to investigate the surface of Earth-like exoplanets by the scattered light curves. We note that such time variations of the scattered light are also applicable to determine the rotation period from as shown by \cite{2008ApJ...676.1319P}.

A variety of inversion techniques of the planetary surface from the scattered light curves have been proposed. Surprisingly, the first theoretical study of scattered light curves to make albedo maps has been performed at the beginning of the twentieth century although the author assume asteroid and satellites for the target \citep{1906ApJ....24....1R}. \citet{2009ApJ...700..915C} performed principal component analysis (PCA) on multi-band photometric data of the Earth observed by EPOXI (Extrasolar Planet Observation and Deep Impact Extended Investigation) mission, and extracted spectral features which roughly correspond to land and ocean. They also checked the time variation of these components and translated it to the longitudinal distribution of these components based on  the formulation by \citet{2008ApJ...678L.129C}. \cite{2009ApJ...700.1428O} paid attention to the gap of reflectivity between ocean and land, and reproduced a longitudinal map of land. 
\cite{Fujii2010} (hereafter F10) have developed a methodology to estimate the areas of ocean, soil, vegetation, and snow from multi-band photometry, and showed that the area of these components can be recovered from mock observations of a cloudless Earth. 

Since these authors focused on diurnal variations in mapping the surface, the resultant maps have only longitudinal resolution with little information of the latitudinal distribution.  One of the goals of the present paper is to develop a method to map inhomogeneous surfaces of exoplanets with both longitudinal and latitudinal resolutions using both diurnal and annual variations of the scattered lights.  

We also consider the determination of the planetary obliquity from time variation of planetary light. The obliquity is an important property of Earth-like planets with its strong implications for climate, habitability \citep[{\it e.g.},][]{1997Icar..129..254W, 2003IJAsB...2....1W} and planetary formation. N-body simulations of the final stage of terrestrial planet formation indicated that the distribution of the obliquity $\zeta$ is isotropic \citep{1999Icar..142..219A,2001Icar..152..205C,2007ApJ...671.2082K}. 
\cite{2004NewA...10...67G} modeled the infrared light curves of exoplanets and showed how the obliquity affects an annual variation. \citet{2009ApJ...700.1428O} also pointed out a possibility to determine the planet's obliquity. In this paper, we demonstrate that the obliquity is estimated simultaneously  with the global map of the planet by analyzing the scattered light curves over its orbital period.

The rest of the paper is organized as follows. We first review the estimation method of the weighted area from multi-band photometry proposed by F10, and describe our methodology to reconstruct the planetary surface and the obliquity measurement in \S 2. Assuming a future satellite mission for the direct imaging of Earth-like planets, we apply our methodology to mock observations based on real data of the scattering properties of the Earth in \S 3. Finally we summarize our results in \S 4.

\section{Methods \label{cp:method}}

\begin{figure*}[!tbh]
  \centerline{\includegraphics[width=150mm]{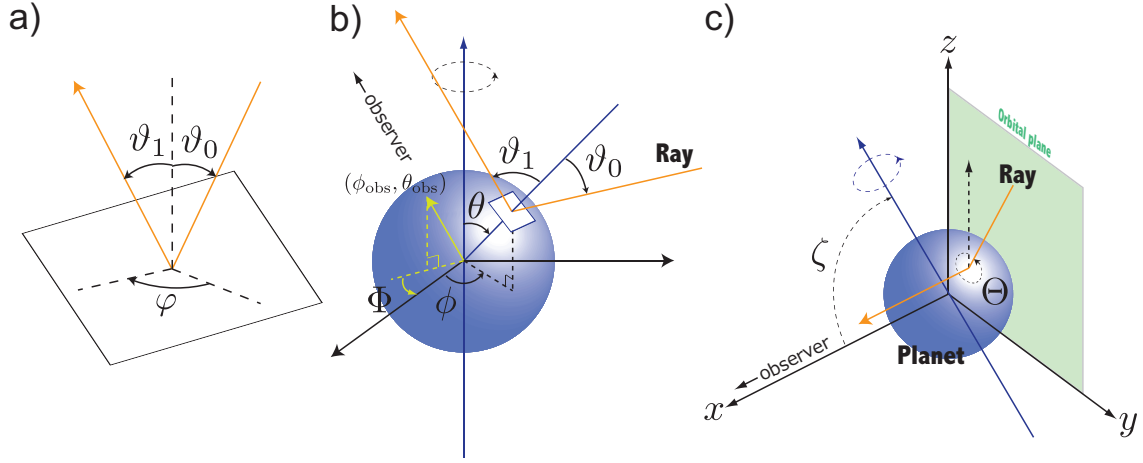}}
  \caption{
  Schematic configurations of the planetary surface and system. Panel a) displays the definitions of the arguments of the BRDF $f(\vartheta_0,\vartheta_1, \varphi ;\lambda)$: the incident zenith angle, $\vartheta_0$, the zenith angle of scattering, $\vartheta_1$, and the relative azimuthal angle  between the incident and scattered light, $\varphi$. 
  Panel b) shows the spherical coordinate fixed on the planetary surface:  the complementary latitude, $\theta$, and the longitude, $\phi$. 
  The spin rotation axis is indicated by $\theta=0$. 
  The spherical coordinate of the vector from the planetary center to observer is denoted by $(\phi_\mathrm{obs},\theta_\mathrm{obs})$. Then, the spin rotation is described by $\Phi \equiv 2 \pi - \phi_\mathrm{obs}$. 
  Panel c) illustrates the coordinate system to specify the phase of the planet. 
   The $\Theta $ denotes the orbital longitude measured from the summer solstice (i.e. the angle between the incident ray and the projected vector of the spin rotation axis to the orbital plane).
 The obliquity $\zeta$ is the angle between the spin rotation axis and a normal vector of the orbital plane. In this paper, we assume that the line of sight is perpendicular to the orbital plane. \label{fig:geo}}
\end{figure*}

Let us briefly summarize the reconstruction method of the planetary surface from scattered light curves by F10. Scattered light from the planetary surface is characterized by the bidirectional reflectance distribution function (BRDF), $f(\vartheta_0,\vartheta_1, \varphi ;\lambda) \mathrm{[str^{-1}]}$, where $\vartheta_0$, $\vartheta_1$, and $\varphi$ are the incident zenith angle, the scattering zenith angle, and the angle between the incident and scattered light, respectively (Fig. [\ref{fig:geo}] a). The BRDF is also a function of the wavelength  $\lambda$. Using the BRDF at $(\phi,\theta)$ fixed on the sphere ( Fig. [\ref{fig:geo}]b), the total scattered intensity at a given phase is provided by 
\begin{eqnarray}
I(\lambda) =  F_\ast (\lambda) R_p^2 \int_{\SVI} f(\phi,\theta; \lambda) \cos{\vartheta_0  (\phi, \theta) } \cos{\vartheta_1  (\phi, \theta)} d \Omega, \nonumber \\
\label{eq:intensity}
\end{eqnarray}
where $F_\ast (\lambda)$ is the incident flux at wavelength $\lambda$, $R_p$ is a planetary radius,  $\SVI$ is the surface area  facing on the observer and illuminated by the host star (we call it the visible and illuminated area), and $d \Omega = \sin \theta d \theta d \phi$. The position on the surface $(\phi,\theta)$ fully specifies $\vartheta_0, \vartheta_1$, and $\varphi$, and thus we write $f(\phi,\theta; \lambda) \equiv f(\vartheta_0 (\phi,\theta),\vartheta_1 (\phi,\theta), \varphi(\phi,\theta) ;\lambda)$.

In order to make a fitting model for the multi-band photometric data, F10 classified the planetary surface into four types: ocean, snow, soil and vegetation. Denoting the specific BRDF of each type $k$ by $f_k (\vartheta_0,\,\vartheta_1,\,\varphi)$, one can write the local BRDF at position $(\phi, \theta)$ as a summation of the specific BRDFs of the $k$-th type surface weighted by local cover fractions $m^{(k)} (\phi, \theta)$:
\begin{eqnarray}
f(\phi,\theta; \lambda) &=& \sum_{k=1}^{N_C} f_k (\phi,\theta; \lambda) m^{(k)} (\phi, \theta),
\end{eqnarray}
where $N_C$ is the number of surface types and $N_C=4$ in this case. Assuming that the scattering is isotropic (Lambertian),
they approximate the BRDF by wavelength-dependence of albedo $a_k(\lambda ) $,
\begin{eqnarray}
 f_k (\phi,\theta; \lambda) = \frac{a_k (\lambda)}{\pi}.
\end{eqnarray}
Under the above assumption, equation (\ref{eq:intensity}) reduces to
\begin{eqnarray}
&\,&I(\lambda) = F_\ast (\lambda) R_p^2 \sum_{k=1}^{N_C} \frac{a_k (\lambda)}{\pi} \nonumber \\ &\times& \int_{\SVI} m^{(k)} (\phi,\theta) \cos{\vartheta_0 (\phi, \theta)} \cos{\vartheta_1 (\phi, \theta)} d \Omega.
\label{eq:le}
\end{eqnarray}

Assuming that they know spectra for each components, F10 developed a reconstruction method for the weighted area of the $k$-th component,
\begin{eqnarray}
A_k &\equiv& R_p^2 \int_{\SVI} W(\phi,\theta) m^{(k)} (\phi,\theta) d \Omega \nonumber \\
W(\phi, \theta) &\equiv& \frac{ \cos{\vartheta_0 (\phi,\theta)}  \cos{\vartheta_1 (\phi,\theta) }}{ \int_{\SVI}  \cos{\vartheta_0 (\phi,\theta)}  \cos{\vartheta_1 (\phi,\theta) } d \Omega},
\label{eq:weia}
\end{eqnarray}
where $W(\phi,\theta)$ is the weight function. 
 The above definition reduces equation (\ref{eq:le}) to a set of linear discrete equation: 
\begin{eqnarray}
C \frac{I(\lambda_b)}{F_\ast (\lambda_b)} &=& \sum_{k=1}^{N_C} a_k (\lambda_b) A_k, \nonumber \\ 
C &\equiv&  \pi \int_{\SVI}  \cos{\vartheta_0 (\phi,\theta)}  \cos{\vartheta_1 (\phi,\theta) } d \Omega ,
\label{eq:invs}
\end{eqnarray}
where $C$ is a factor that depends on the phase angle, {\it i.e.}, the planetocentric angle between the host star and the observer, and $\lambda_b$ is the center wavelength of the $b$-th band.  Using a cloudless model Earth, they showed that the weighted area can be approximately recovered by solving equation (\ref{eq:invs}) with an inequality condition $A_k > 0$, and the summation of $A_k(t)$ is approximately constant over time, $\sum A_k (t) \approx const. \approx R_p^2$. It means that one can roughly estimate $R_p$ as well by their method. Therefore, we use the fractional area $A_k/R_p^2$ normalized by $R_p^2$ in what follows. F10 have considered the spin rotation of a planet only and recovered the weighted area during one day of the planet and the weighted area can be converted to longitudinal maps by the inversion method proposed by \citet{2008ApJ...678L.129C} and \citet{2009ApJ...700..915C}. 

\subsection{An Inverse Problem of a Two-dimensional Map}

In the present paper, we consider the orbital motion to recover a two-dimensional world map of the planetary surface. We assume a model planet on a face-on circular orbit, which is the most promising case for the planet mapping. We also assume that orbital period is $P_{orb}=365$ [day] and spin rotational period is $P_{spin}=24$ [hrs]. 
We denote the phases of the orbital motion and spin rotation by $\Theta$ and $\Phi$ (Fig. [\ref{fig:geo}] c), respectively.  Our mock observation is assumed to be performed over one year from $\Theta=\Theta_S$ to $\Theta = \Theta_S + 2 \pi$. 
There are two unknown parameters in this geometry: the obliquity $\zeta$ and the initial orbital longitude $\Theta_S$. We discuss the measurement of $\zeta$ and $\Theta_S$ in section \ref{cp:zeta}. 

For our conventional geometry, the weight function is given by
\begin{eqnarray}
&\,& W(\phi,\theta;\Theta;\Phi;\zeta) = \frac{3}{2} \cos{\vartheta_0} \cos{\vartheta_1} \nonumber \\
&=& \frac{3}{2} [ \, \cos{\theta} \cos{\Theta} \sin{\zeta} - \sin{\theta} \sin{(\phi+\Phi)} \sin{\Theta} \nonumber \\ &-&  \sin{\theta} \cos{(\phi+\Phi)} \cos{\Theta} \cos{\zeta} \, ] \nonumber \\ 
&\times& \left[ \sin{\theta} \cos{(\phi+\Phi)}  \sin{\zeta}  +\cos{\theta} \cos{\zeta} \right].
\end{eqnarray}
We provide the explicit form of the visible and illuminated area, $\SVI$ in Appendix A. Note that the weight function is normalized to unity by integrating over the sphere:
\begin{eqnarray}
 \int W(\phi,\theta;\Theta;\Phi;\zeta) d \Omega = 1.
\end{eqnarray}

\begin{figure*}[!tbh]
\includegraphics[width=85mm]{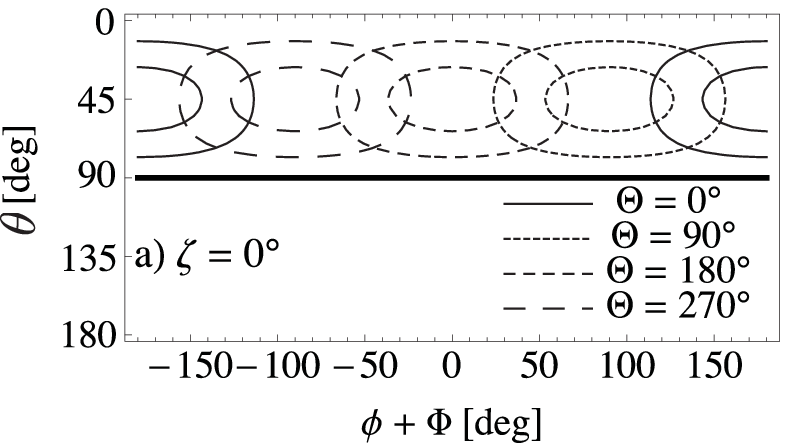}
\includegraphics[width=85mm]{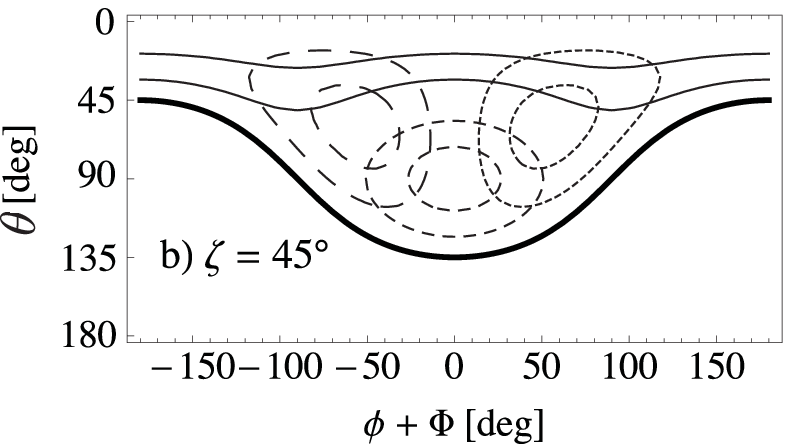}
\includegraphics[width=85mm]{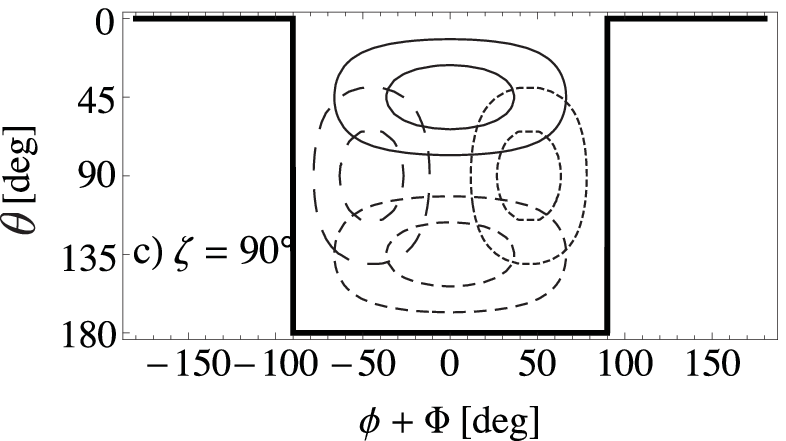}
  \caption{Geometry dependence of the weight function $W(\phi,\theta;\Theta;\Phi;\zeta)$ in the visible and illuminated area $\SVI$. Annual variations are indicated by different thin lines. Solid, large-dashed, middle-dashed and small-dashed lines indicate $\Theta=0^\circ, 90^\circ, 180^\circ, $ and $270^\circ$, respectively. Because the weight function solely depends on $\phi + \Phi$, in other words, longitude minus the phase of the spin rotation, we adopt $\phi + \Phi$ to $x$-axes instead of showing diurnal variations explicitly. } Thick lines indicate the boundary of the visible area, $\theta = \theta_V  (\phi;\Phi;\zeta)$ in equation (\ref{eq:g}). Each panel shows a different obliquity, $\zeta=0^\circ$ (panel a), $45^\circ$ (panel b), and $90^\circ$ (panel c).  Inner and outer contours indicate $W(\phi,\theta;\Theta;\Phi;\zeta)=0.6$ and $0.3$, respectively. \label{fig:gref}
\end{figure*}

Figure \ref{fig:gref} shows the behavior of the weight function $W(\phi,\theta;\Theta;\Phi;\zeta)$ for cases of $\zeta =0^\circ$, $45^\circ$ and $90^\circ$. The weight function covers the region of $0^\circ < \theta < 90^\circ + \zeta$ and $0^\circ < \phi < 360^\circ $ as the planet rotates around the host star and its spin axis. As a result, the scattered light curves contain information about the land distribution  up to $ 50 \, (1 + \sin{\zeta}) \% $ of the total surface area in principle. 

We consider how to deduce the local cover fraction $m^{(k)} (\phi,\theta)$ if a set of the weighted area throughout the orbital and spin rotational periods is given. In order to solve this problem, we apply one of the techniques of tomography, the linear inverse problem, to the data. We assume that the planetary surface consists entirely of land and ocean and do not consider the other components nor further decompose them into clouds, vegetation or soil in what follows. Therefore we use the land cover fraction $m(\phi,\theta)$ for the surface classification ($k= \mathrm{land}$). The surface at $(\phi,\theta)$ covered by land only (ocean only) is expressed by $m(\phi,\theta)=1$ ($m(\phi,\theta)=0$). We also denote the weighted area of land by $A(\Phi(t_i),\Theta(t_i);\zeta)$.  In \S 3, we also consider the components of vegetation and soil using mock photometric data.

To begin with, let us discretize the observing time in $\Ndata$ epochs and pixelize the planetary surface in $\Mmodel$ pixels. The weighted area recovered at the $i$-th epoch $t_i$ is written as
\begin{eqnarray}
&A&(\Phi(t_i),\Theta(t_i);\zeta) \nonumber \\ &=& \sum_{j \,|\,(s_j \, \cap \, \SVI) \neq \emptyset } \, \int_{s_j} W(\phi,\theta;\Theta(t_i);\Phi(t_i);\zeta) m(\phi,\theta) d \Omega \nonumber \\
&=&  \sum_{j \,|\,(s_j \, \cap \, \SVI) \neq \emptyset }  \, \langle m \rangle_{ij} \int_{s_j} W(\phi,\theta;\Theta(t_i);\Phi(t_i);\zeta) d \Omega,
\label{eq:linialization1}
\end{eqnarray}
where $s_j$ indicates the $j$-th pixel on the surface. The weighted cover fraction $\langle m \rangle_{ij}$ is defined as
\begin{eqnarray}
 \langle m \rangle_{ij} \equiv \frac{\int_{s_j} W(\phi,\theta;\Theta(t_i);\Phi(t_i);\zeta) m(\phi,\theta) d \Omega}{\int_{s_j} W(\phi,\theta;\Theta(t_i);\Phi(t_i);\zeta) d \Omega }.
\end{eqnarray} 
Here, let us define the pixel-averaged cover fraction
\begin{eqnarray}
\overline{m}_j  \equiv  \frac{\int_{s_j} m(\phi,\theta) d \Omega}{\int_{s_j}  d \Omega}.
\end{eqnarray}
Under the assumption that the weighted cover fraction approximately equals to the pixel-averaged cover fraction:
\begin{eqnarray}
  \langle m \rangle_{ij} \approx \overline{m}_j, 
\end{eqnarray}
we obtain 
\begin{eqnarray}
A(\Phi(t_i),\Theta(t_i);\zeta) &=& \sum_j \overline{m}_j \int_{s_j} W(\phi,\theta;\Theta(t_i);\Phi(t_i);\zeta) d \Omega. \nonumber \\
\label{eq:lp1}
\end{eqnarray}
We introduce the data vector of weighted area $\A$, the design matrix $G$ and the model vector of cover fraction $\m$ as
\begin{eqnarray}
\A &\equiv& \{a_i = A(\Phi(t_i),\Theta(t_i);\zeta)/\sigma_i \,| \, i=1,2,...,\Ndata \}, \nonumber \\
G &\equiv& \{ G_{ij} = \int_{s_j} W(\phi,\theta;\Theta(t_i);\Phi(t_i);\zeta) d \Omega/\sigma_i \, \nonumber \\ &|& \, i=1,2,...,\Ndata, j=1,2,...,\Mmodel   \},  \nonumber \\
\m &\equiv& \{\overline{m}_j \,| \, j=1,2,...,\Mmodel \},
\label{eq:defagm}
\end{eqnarray}
where $\sigma_i$ is the error of the weighted area. 
Then we can rewrite equation (\ref{eq:lp1}) as
\begin{eqnarray}
\A &=& G \, \m.
\label{eq:lip}
\end{eqnarray}
The above equation can be solved for $\m$ by minimizing $\chi^2$:
\begin{eqnarray}
\chi^2 \equiv |\Adata - G \, \m|^2 ,
\label{eq:least_square}
\end{eqnarray}
under inequality constraints:
\begin{eqnarray}
0 \le m_{j} \le 1,
\label{eq:cond}
\end{eqnarray}
where $\Adata$ is the observed weighted area with noise. We denote $\m$ that minimizes equation (\ref{eq:least_square}) by $\mest$.

\subsection{Solving the Inverse Problem}

\begin{figure*}[!tbh]
      \includegraphics[width=75mm]{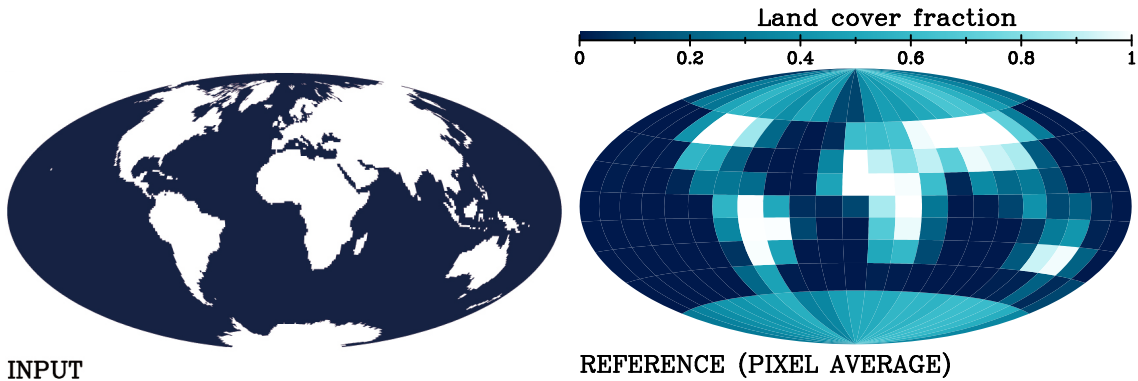}
      \includegraphics[width=75mm]{pixel_average.ps}
 \caption{The input map (left) and the reference map (right). The left panel shows the land distribution of the Earth based on the data set of the ISLSCP II fixed land/water masks (http://islscp2.sesda.com). Right panel indicates the pixel-averaged land cover fraction ($\overline{m}_j$) computed by averaging the left panel over pixels. \label{fig:in}}
\end{figure*}

We are now in a position to solve equation (\ref{eq:least_square}) with the condition (\ref{eq:cond}) using an idealized data set. Since one can choose an arbitrary division for the surface modeling, we use the following pixelization:
\begin{eqnarray}
s_{j} &=& \{(\phi,\theta) | 1 - (1+\sin{\zeta}) \frac{j_\theta}{M_\theta}  < \cos{\theta} \le \nonumber \\ &\,& 1 - (1+\sin{\zeta}) \frac{j_\theta - 1}{M_\theta},  \frac{2 \pi}{M_\phi} (j_\phi-1) \le \phi < \frac{2 \pi}{M_\phi} j_\phi \}, \nonumber \\
j &=&  (j_\phi - 1) M_\theta + j_\theta \nonumber \\
&\,& \mbox{ for $j_\phi = 1,2,,,M_\phi$ and $j_\theta = 1,2,,,M_\theta$}.
\end{eqnarray}
The above pixelization has a constant solid angle for a given $\zeta$ and the number of pixels does not change for different $\zeta$'s. We adopt $M_\phi=9$ and $M_\theta =  23$ (see the right panel of Figure \ref{fig:in} for the $\zeta = 90^\circ$ case) and assume the numbers of epochs for diurnal variations $N_\Phi = 23$ and for annual variations $N_\Theta =23$.  In Appendix B, we discuss the dependence of model degeneracies on the pixelization. 

We create the synthetic data of the weighted area using $1^\circ \times 1^\circ$ fixed land/water masks of the ISLSCP II ({\it input map}; left panel of Figure \ref{fig:in}). Pixel averaged land cover fraction of this data ({\it reference map}) is shown in the right panel of Figure \ref{fig:in}.
We fix the obliquity $\zeta=90^\circ$ and $\Theta(t_0) = \Theta_S=0^\circ$ and compute the weighted area $A_\mathrm{input} (\Phi, \Theta; \zeta=90^\circ)$ from the input map (not from the reference map). We compute the data $A_\mathrm{data}$ by adding a Gaussian noise $N_g$ to $A_\mathrm{input} (\Phi, \Theta; \zeta=90^\circ)$:
\begin{eqnarray}
A_\mathrm{data} (\Phi, \Theta) = A_\mathrm{input} (\Phi, \Theta; \zeta=90^\circ) +  N_g.
\end{eqnarray}
We denote the standard deviation of the Gaussian random variable $N_g$ by $\sigma \equiv \langle N_g^2 \rangle^{1/2}$.

 The linear inverse problem (eq. [\ref{eq:lip}]), or the equivalent least square problem (eq. [\ref{eq:least_square}]), under the inequality constraints (eq. [\ref{eq:cond}]) can be solved with Bounded Variable Least Squares Solver (BVLS) developed by \cite{1974slsp.book.....L,1995LH} \footnote[1]{The original code of the BVLS is available through NETLIB (http://www.netlib.org/lawson-hanson/index.html)}. The BVLS is a generalization of the non-negative least square (NNLS) described in  \cite{1974slsp.book.....L,1995LH} and uses the QR decomposition \citep[see also ][]{1989gdad.book.....M}. The BVLS works for both overdetermined and underdetermined (ill-condition) problems. Even for ill-condition problem, the BVLS provides one of the (non-unique) solutions \citep{1995LH}. 

\begin{figure*}[]
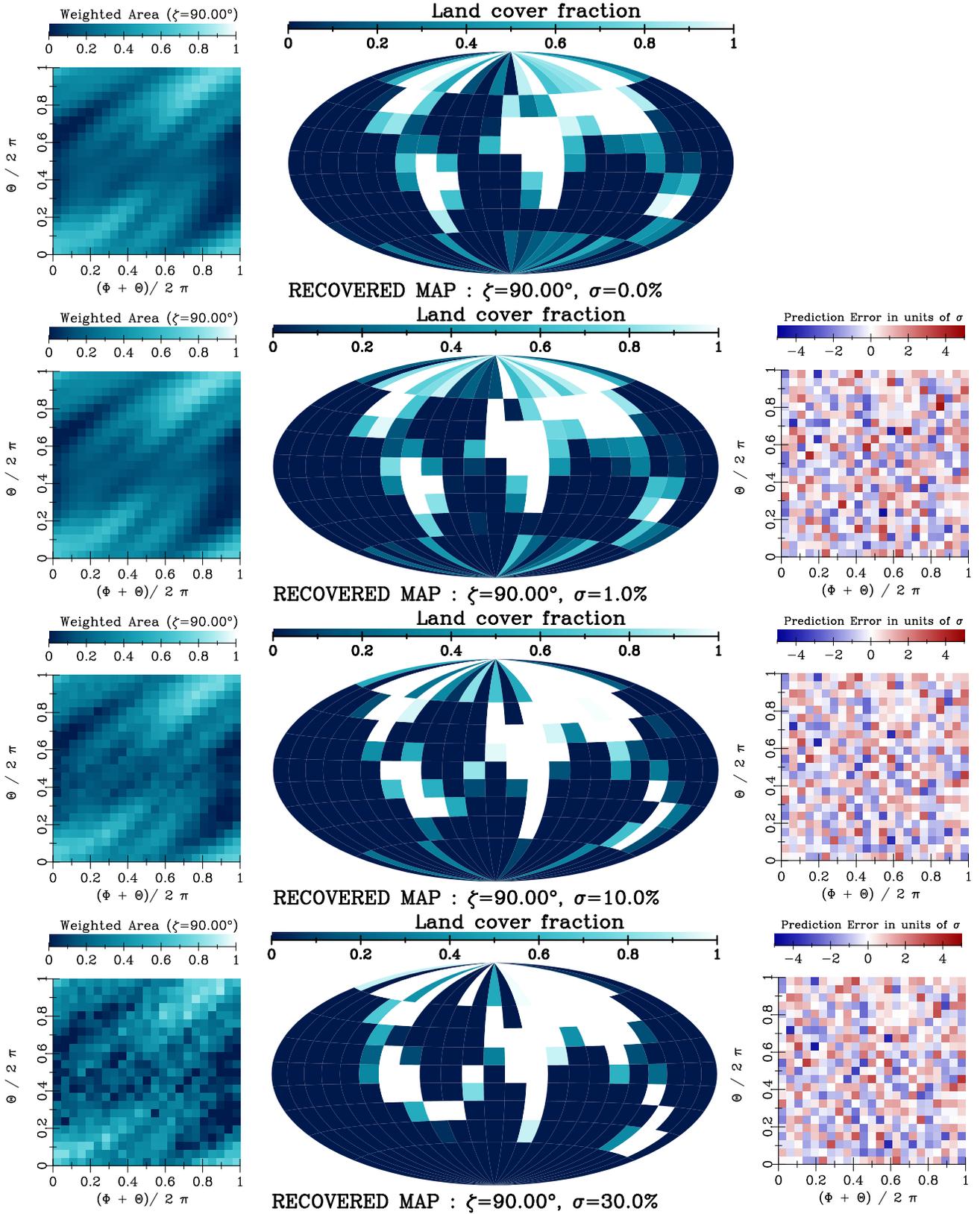

  \begin{minipage}{0.25\linewidth}
    \begin{center}
      \includegraphics[width=\linewidth]{amap0.0-90.00.ps}
    \end{center}
  \end{minipage}
  \begin{minipage}{0.45\linewidth}
    \begin{center}
      \includegraphics[width=\linewidth]{recover_extreme0.0-90.00.ps}
    \end{center}
  \end{minipage}
  \begin{minipage}{0.25\linewidth}
    \begin{center}
      \vspace{2cm}
    \end{center}
  \end{minipage}
  \begin{minipage}{0.25\linewidth}
    \begin{center}
      \includegraphics[width=\linewidth]{amap1.0-90.00.ps}
    \end{center}
  \end{minipage}
  \begin{minipage}{0.45\linewidth}
    \begin{center}
      \includegraphics[width=\linewidth]{recover_extreme1.0-90.00.ps}
    \end{center}
  \end{minipage}
  \begin{minipage}{0.25\linewidth}
    \begin{center}
      \includegraphics[width=\linewidth]{amap_preerr1.0-90.00.ps}
    \end{center}
  \end{minipage}
  \begin{minipage}{0.02\linewidth}
    \begin{center}
      \vspace{2cm}
    \end{center}
  \end{minipage}
  \begin{minipage}{0.25\linewidth}
    \begin{center}
      \includegraphics[width=\linewidth]{amap10.0-90.00.ps}
    \end{center}
  \end{minipage}
  \begin{minipage}{0.45\linewidth}
    \begin{center}
      \includegraphics[width=\linewidth]{recover_extreme10.0-90.00.ps}
    \end{center}
  \end{minipage}
  \begin{minipage}{0.25\linewidth}
    \begin{center}
      \includegraphics[width=\linewidth]{amap_preerr10.0-90.00.ps}
    \end{center}
  \end{minipage}
  \begin{minipage}{0.02\linewidth}
    \begin{center}
      \vspace{0.5cm}
    \end{center}
  \end{minipage}
  \begin{minipage}{0.25\linewidth}
    \begin{center}
      \includegraphics[width=\linewidth]{amap30.0-90.00.ps}
    \end{center}
  \end{minipage}
  \begin{minipage}{0.45\linewidth}
    \begin{center}
      \includegraphics[width=\linewidth]{recover_extreme30.0-90.00.ps}
    \end{center}
  \end{minipage}
  \begin{minipage}{0.25\linewidth}
    \begin{center}
      \includegraphics[width=\linewidth]{amap_preerr30.0-90.00.ps}
    \end{center}
  \end{minipage}
  \begin{minipage}{0.02\linewidth}
    \begin{center}
      \vspace{0.5cm}
    \end{center}
  \end{minipage}
 \caption{Two dimensional images of the weighted area $A(\phi,\theta)$ (left panels), recovered maps (middle panels) for $\zeta=90^\circ$, and prediction errors, $[A_\mathrm{data} (\Phi(t_i), \Theta(t_i)) - G \mest]/\sigma$ for different noises, $\sigma=0, 0.01  \langle A_\mathrm{input} \rangle, 0.1 \langle A_\mathrm{input} \rangle,$ and $ 0.3 \langle A_\mathrm{input} \rangle$ from the top to the bottom panels.   \label{fig:extreme}}
\end{figure*}

By solving the least square problem (eq. [\ref{eq:least_square}]) with the BVLS algorithm, we obtain the recovered map $\mest$. Figure \ref{fig:extreme}  displays the two-dimensional images of the weighted area (left), the recovered maps (middle), and the prediction errors, $[A_\mathrm{data} (\Phi(t_i), \Theta(t_i)) - G \mest]/\sigma $ (right) for $\zetain=90^\circ$. In this figure, we regard $\zeta$ and $\Theta_S$ as fixed parameters and use the input values $\zeta = \zetain = 90^\circ$ and $\Theta_S = \ThetaSin = 0^\circ$  for equation (\ref{eq:defagm}). We consider four cases of the noise, $\sigma=0, 0.01  \langle A_\mathrm{input} \rangle, 0.1 \langle A_\mathrm{input} \rangle,$ and $ 0.3 \langle A_\mathrm{input} \rangle$, where $\langle A_\mathrm{input} \rangle$ is the average of $A_\mathrm{input} (\phi,\theta; \zeta=90^\circ)$ and is approximately equal to $0.3$ for the case of the Earth.  As discussed in Appendix C, we confirmed that the estimated map is the unique solution under the inequality constraint. Comparing the recovered maps with the reference map (right panel in Fig. \ref{fig:in}), we conclude that one can approximately recover the land distribution from the two-dimensional map of the weighted area.

\begin{figure*}[!tbh]
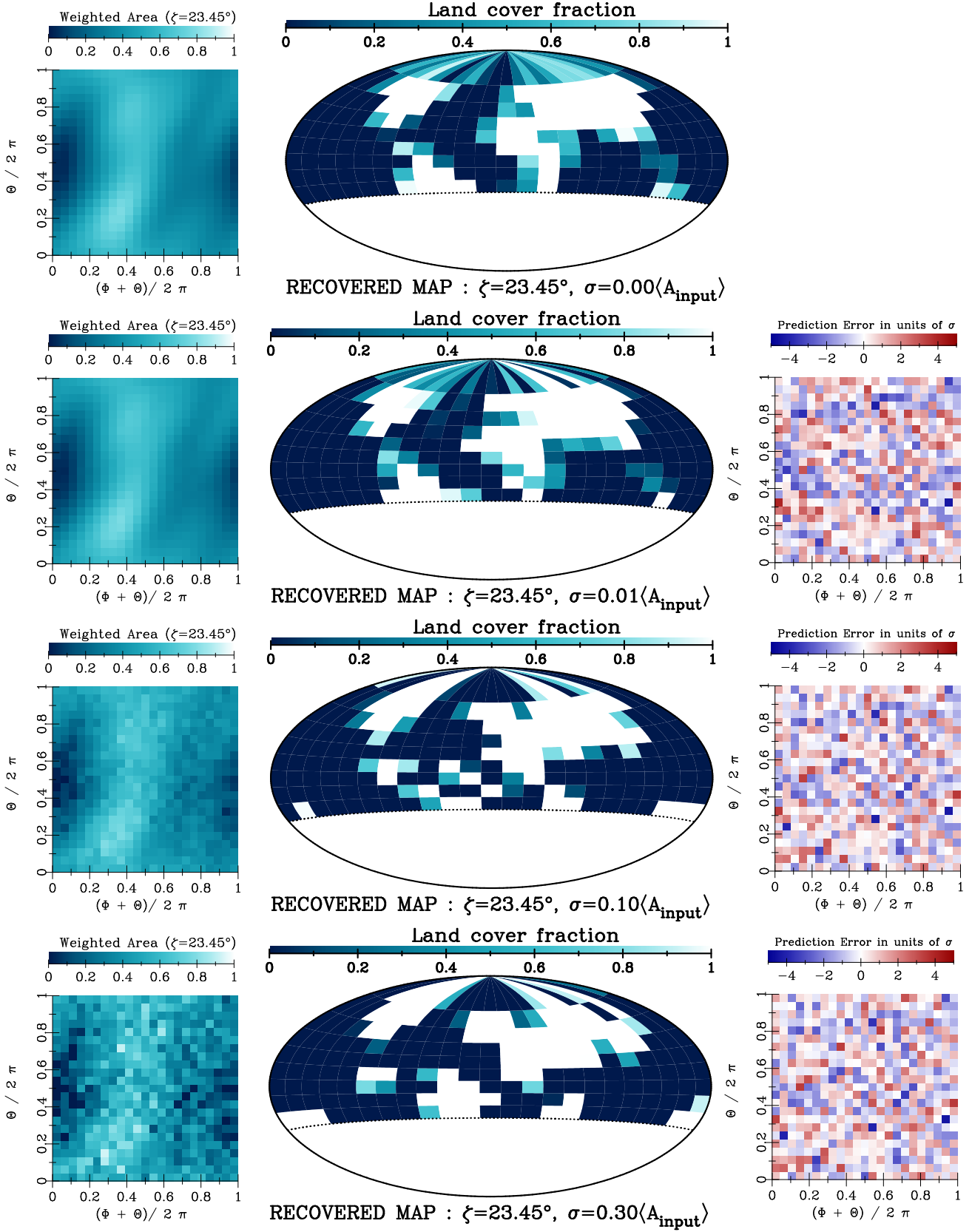

  \begin{minipage}{0.25\linewidth}
    \begin{center}
      \includegraphics[width=\linewidth]{amapvar0.0-23.45.ps}
    \end{center}
  \end{minipage}
  \begin{minipage}{0.45\linewidth}
    \begin{center}
      \includegraphics[width=\linewidth]{recover_varextreme0.0-23.45.ps}
    \end{center}
  \end{minipage}
  \begin{minipage}{0.25\linewidth}
    \begin{center}
      \vspace{2cm}
    \end{center}
  \end{minipage}
  \begin{minipage}{0.25\linewidth}
    \begin{center}
      \includegraphics[width=\linewidth]{amapvar1.0-23.45.ps}
    \end{center}
  \end{minipage}
  \begin{minipage}{0.45\linewidth}
    \begin{center}
      \includegraphics[width=\linewidth]{recover_varextreme1.0-23.45.ps}
    \end{center}
  \end{minipage}
  \begin{minipage}{0.25\linewidth}
    \begin{center}
      \includegraphics[width=\linewidth]{amapvar_preerr1.0-23.45.ps}
    \end{center}
  \end{minipage}
  \begin{minipage}{0.02\linewidth}
    \begin{center}
      \vspace{2cm}
    \end{center}
  \end{minipage}
  \begin{minipage}{0.25\linewidth}
    \begin{center}
      \includegraphics[width=\linewidth]{amapvar10.0-23.45.ps}
    \end{center}
  \end{minipage}
  \begin{minipage}{0.45\linewidth}
    \begin{center}
      \includegraphics[width=\linewidth]{recover_varextreme10.0-23.45.ps}
    \end{center}
  \end{minipage}
  \begin{minipage}{0.25\linewidth}
    \begin{center}
      \includegraphics[width=\linewidth]{amapvar_preerr10.0-23.45.ps}
    \end{center}
  \end{minipage}
  \begin{minipage}{0.02\linewidth}
    \begin{center}
      \vspace{0.5cm}
    \end{center}
  \end{minipage}
  \begin{minipage}{0.25\linewidth}
    \begin{center}
      \includegraphics[width=\linewidth]{amapvar30.0-23.45.ps}
    \end{center}
  \end{minipage}
  \begin{minipage}{0.45\linewidth}
    \begin{center}
      \includegraphics[width=\linewidth]{recover_varextreme30.0-23.45.ps}
    \end{center}
  \end{minipage}
  \begin{minipage}{0.25\linewidth}
    \begin{center}
      \includegraphics[width=\linewidth]{amapvar_preerr30.0-23.45.ps}
    \end{center}
  \end{minipage}
  \begin{minipage}{0.02\linewidth}
    \begin{center}
      \vspace{0.5cm}
    \end{center}
  \end{minipage}
 \caption{Two dimensional images of the weighted area $A(\phi,\theta)$ (left panels) and recovered maps (right panels)  for $\zeta=23.45^\circ $. \label{fig:extreme2345}}
\end{figure*}

We also solve the inverse problem in the case of the obliquity of the Earth, $\zetain=23.45^\circ$ (Figure \ref{fig:extreme2345}). Even for a small obliquity like the Earth, one can recover the land distribution of $50 \, [1 + \sin{(\zeta=23.45^\circ)}] \approx 70$ \% of the planetary surface. We conclude that our methodology can recover main features of the continents on the planetary surface even if the planet has small obliquity as is the case of the Earth.

\subsection{Measurement of Obliquity \label{cp:zeta}}

\begin{figure*}[!tbh]
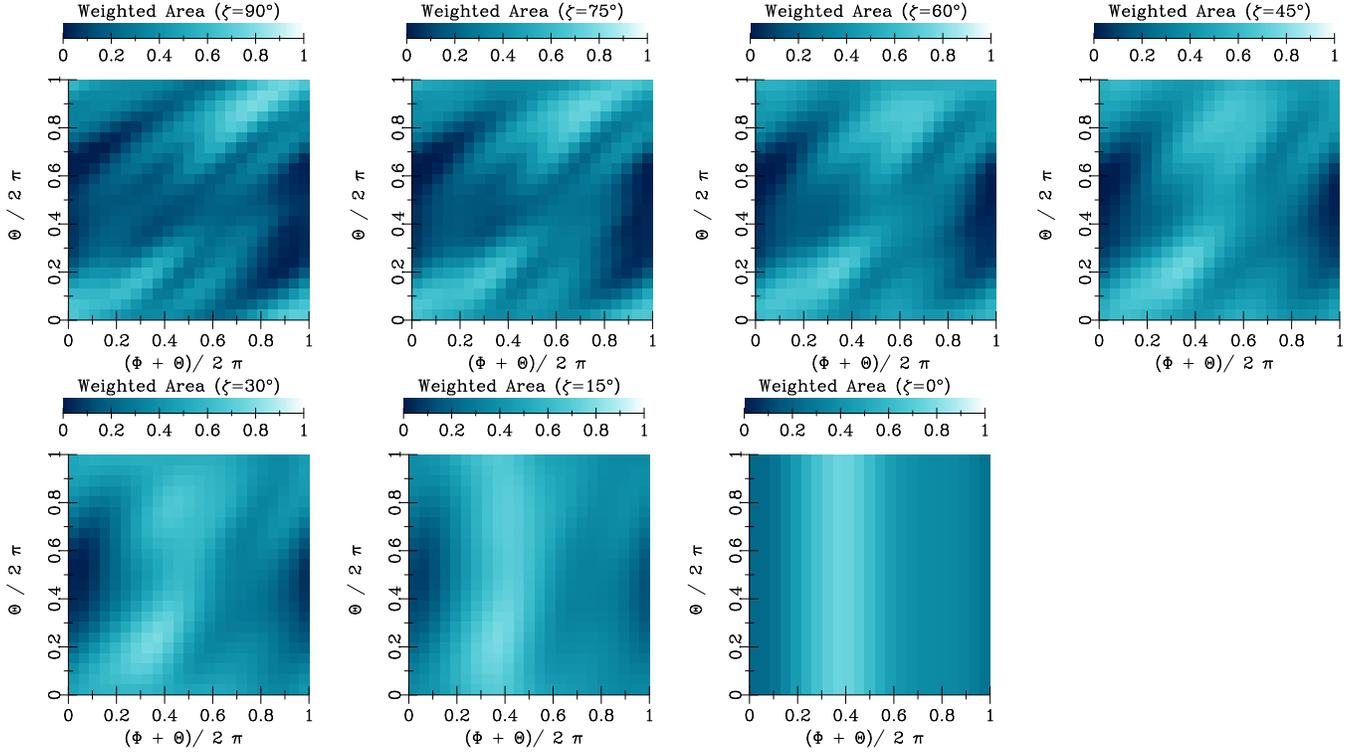

  \begin{minipage}{0.24\linewidth}
    \begin{center}
      \includegraphics[width=\linewidth]{amap0.0-90.ps}
    \end{center}
  \end{minipage}
  \begin{minipage}{0.24\linewidth}
    \begin{center}
      \includegraphics[width=\linewidth]{amap0.0-75.ps}
    \end{center}
  \end{minipage}
  \begin{minipage}{0.24\linewidth}
    \begin{center}
      \includegraphics[width=\linewidth]{amap0.0-60.ps}
    \end{center}
  \end{minipage}
  \begin{minipage}{0.24\linewidth}
    \begin{center}
      \includegraphics[width=\linewidth]{amap0.0-45.ps}
    \end{center}
  \end{minipage}
  \begin{minipage}{0.24\linewidth}
    \begin{center}
      \includegraphics[width=\linewidth]{amap0.0-30.ps}
    \end{center}
  \end{minipage}
  \begin{minipage}{0.24\linewidth}
    \begin{center}
      \includegraphics[width=\linewidth]{amap0.0-15.ps}
    \end{center}
  \end{minipage}
  \begin{minipage}{0.24\linewidth}
    \begin{center}
      \includegraphics[width=\linewidth]{amap0.0-0.ps}
    \end{center}
  \end{minipage}
 \caption{Obliquity dependence of two dimensional images of the weighted area. \label{fig:oblidep}}
\end{figure*}

\begin{figure*}[!tbh]
      \includegraphics[width=60mm]{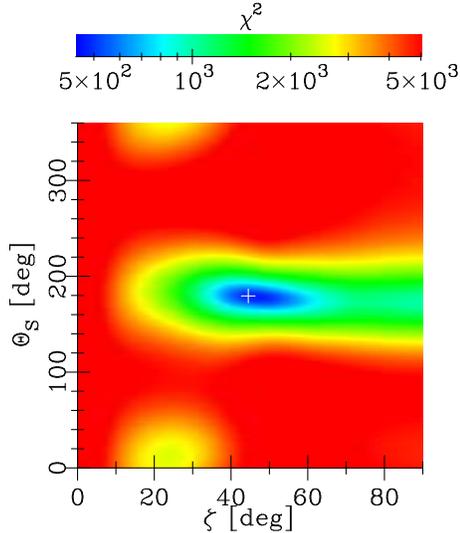}
      \caption{$\chi^2(\zeta,\Theta_S)$ for $\zetain = 45^\circ $ and $ \ThetaSin = 180^\circ$ (cross) with 10 \% additional noise. The estimated obliquity and initial orbital longitude are  $\zetaest = 46.3^\circ \pm 0.6^\circ $  and $\ThetaSest= 181.8^\circ \pm 0.8^\circ$, respectively. \label{fig:chisq_zt}}
\end{figure*}

So far, we have assumed that the obliquity $\zeta$ and the initial phase $\Theta_S$ in the orbit are known a priori.  Next we try to estimate the obliquity $\zeta$ by our method itself. Figure \ref{fig:oblidep} indicates the obliquity dependence of images of the weighted area. We determine the best-fit value of $\zeta$ and $\Theta_S$ by minimizing $\chi^2$ distance:
\begin{eqnarray} 
\chi^2 (\zeta, \Theta_S) &\equiv& \left| \A - G(\zeta,\Theta_S) \mestzt \right|^2 \nonumber \\
G(\zeta,\Theta_S) &\equiv& \{ G_{ij} (\zeta,\Theta_S) = \int_{s_j} W(\phi,\theta;\Theta(t_i);\Phi(t_i);\zeta) d \Omega/\sigma_i \nonumber \\
&\,& | \, i=1,2,...,\Ndata, j=1,2,...,\Mmodel   \} \nonumber, \\
\Theta(t_0) &=& \Theta_S,
\label{eq:chi2}
\end{eqnarray} 
 where $\mestzt$ is determined by minimizing $\chi^2$ with given $\zeta$ and $\Theta_S$ using the BVLS fitting. Figure \ref{fig:chisq_zt} displays $\chi^2 (\zeta, \Theta_S) $ for $\zetain = 45^\circ $ and $ \ThetaSin = 180^\circ$. We use the {\it amoeba} routine \citep{1992nrfa.book.....P} to search for $\zeta$ and $\Theta_S$, which minimizes $\chi^2 (\zeta, \Theta_S)$. We estimate errors of $\zeta$ and $\Theta_S$ by the bootstrap resampling because errors are originated not only from additional Gaussian noise but also from pixelization, linearization, and other systematics. After $10^3$ iterations of the bootstrap, we obtain $\zetaest = 46.3^\circ \pm 0.6^\circ $  and $\ThetaSest= 181.8^\circ \pm 0.8^\circ$  for 10 \% additional noise and  $\zetaest = 50.3^\circ \pm 2.3^\circ $  and $\ThetaSest= 183.0^\circ \pm 2.3^\circ$  for 30 \% additional noise. The estimated values of both obliquity and initial phase in orbit remarkably agree well with the input values.

\section{Applying to Mock Observation of Multi-band Photometry  \label{cp:comp}}

Now, we apply our methodology to a mock multi-band observation of the cloudless Earth as a more realistic demonstration. 
We simulate scattered light curves of a cloudless Earth in the same manner as the simulation in F10. 
We use the solar spectrum as an incident flux on the model Earth. 
We assume an observer seeing the planet on a face-on and circular orbit at a distance of 5 pc (Case A) or 10 pc (Case B) from it.
The scattered light from land is computed using the BRDF of actual land surface assigned by MODIS data-set (``snow-free gap-filled MODIS BRDF Model Parameters''). 
We adopt the data of April and ignore the seasonal variation of the BRDF at each point on the planetary surface. 
The scattered light from ocean is calculated with the BRDF model for wavy ocean described in \citet{nakajima1983}.  
In our model, the snow cover regions around the poles are replaced by ocean. 
We also include the effect of Rayleigh scattering by atmosphere with single scattering approximation between the atmosphere and the underlying surface. 
Neither the effect of clouds nor the molecular absorption is, however, incorporated.

\begin{table*}[!tbh]
\begin{center}
\caption{Observation parameters \label{tab:conf2}}
  \begin{tabular}{cccc}
   \hline\hline
Symbol	&	Quantity & Value & Unit	\\ \hline 
$t_{\rm exp}$	&	exposure time & 24/23$\times$3600 &	sec  \\ 
$n$	&	folded days  & 14 &		days \\ 
$D$	&	diameter of telescope aperture	& 1.1 &	m	\\ 
$\Psi $	&	sharpness	&	0.0433	&		\\
$\alpha	$	&	pixel scale	&	0.03125 &	arcsec/pixel \\
$h$	&	end-to-end efficiency &	0.5	&  \\
$\upsilon$	&	dark rate 	& 0.001	&	counts/sec	\\
$\kappa$		&	read noise 	& 2 		&	$\sqrt{{\rm counts}}$/read	\\
QE	&	quantum efficiency	 & 0.91	&	\\
$\Omega _z$	&	zodiacal light in magnitude &		23	&	mag/square arcsec	\\
$Z$	&	zodiacal light  &	1	&	\\
$F_0$	&	zero flux	& $1.4 \times 10^4$ (band 1) & $ \mathrm{cts/cm^2/nm/s}$ \\
        &		& $9.6 \times 10^3$(band 2)  & $ \mathrm{cts/cm^2/nm/s}$ \\
        &		& $7.2 \times 10^3$(band 3)  & $ \mathrm{cts/cm^2/nm/s}$ \\
        &		& $4.1 \times 10^3$(band 4)  & $ \mathrm{cts/cm^2/nm/s}$ \\
\hline
\end{tabular}
\end{center}
\end{table*}

\begin{table}[!tbh]
\begin{center}
\caption{Band definition. \label{tab:band}}
  \begin{tabular}{ccc}
   \hline\hline
band & wavelength & band width \\ 
	&$\lambda_b $ [$\mu $m]	& $\Delta \lambda $ [$\mu $m]\\ \hline
1 & 0.469 &	0.1	\\
2 & 0.555 &	0.1	\\
3 & 0.645 &	0.1	\\
4 & 0.8585 &	0.1	\\
\hline
\end{tabular}
\end{center}
\end{table}

Our assumed observation parameters are listed in Table \ref{tab:conf2}. 
As an observing system, we assume a satellite telescope with 1.1 m aperture. This assumption comes from the basic architecture proposed for {\it the occulting ozone observatory} (O3) \citep{kasdin2010}, which is a satellite mission for direct imaging of exoplanets in UV/optical/near-IR bands shading the light from the host star by a 30 m external occulter. 
The scattered light is computed at the central wavelength of each band and multiplied by its band width.  
We consider four bands centered at the wavelengths listed in Table \ref{tab:band}, which correspond to the bands of the MODIS land BRDF data. We assume $0.1\,\mu m$ as the bandwidth.

 We divide the orbital phase into 23 ($\Theta_i = 2\pi i/23,\;i=0, 1, ..., 22$), and the spin rotational phase into 23 ($\Phi_i = 2\pi i/23,\;i=0, 1, ..., 22$). We compute the scattered light curves in 4 bands with the exposure time $t_{\rm exp} = 24/23$ hours throughout one year.  Then, we stack data for 14 days centered at each orbital phase $\Theta_i$, and fold the light curves according to its spin rotation period to obtain diurnal curves at the orbital phase, assuming that the spin rotational period is already known through periodogram analysis \citep{2008ApJ...676.1319P}. 
Thus, the resultant light curves have $23 \times 23$ data points for each band. 

As for the noise, we consider the photon shot noise, the exzodiacal light, dark noise, and read noise. 
We ignore the leakage of the light from the host star. 
Thus, the signal to noise ratio S/N is expressed as
\begin{equation}
{\rm S/N} = \frac{S}{\sqrt{S+N_{\rm z}+N_{\rm d}+N_{\rm r}}}, 
\end{equation}
where $S$, $N_{\rm z}$, $N_{\rm d}$, and $N_{\rm r}$ are the scattered light from the planet, the contribution from zodiacal light, dark noise, and read noise, respectively. 
Specifically, they are given as   
\begin{eqnarray}
N_{\rm z} &\equiv& hZF_0 \alpha ^2 10^{-\frac{\Omega _z}{2.5}} \frac{1}{\Psi}\,\pi \left( \frac{D}{2} \right)^2 t_{\rm exp} n \Delta \lambda \\
N_{\rm d} &\equiv& \frac{\upsilon t_{\rm exp} n}{\Psi }\\
N_{\rm r} &\equiv& \frac{\kappa ^2 n}{\Psi }.
\end{eqnarray}
The parameters and their values we adopt are listed in Table \ref{tab:conf2} \citep[see e.g.,][]{2010PASP..122..401S}. 
The zero flux $F_0$ for calibration is calculated by linearly interpolating the spectrum of Vega given by Tables 1 and 2 in \citet{colina1996}, and shown in Table \ref{tab:noise} in this paper. 
In order to quantitatively assess each noise level, we list the amplitudes of these noises per epoch in Table \ref{tab:noise} together with the total average of the signal $S$ in the case of $\zeta = 90 ^{\circ}$. The averaged signal-to-noise per epoch is also listed in Table \ref{tab:noise}. We note that the averaged signal per epoch is comparable with the noise level.

\begin{table*}[!tbh]
\begin{center}
\caption{Assumed signal and noises in 4 bands. \label{tab:noise}}
  \begin{tabular}{lccccc}
   \hline\hline
band $b$ & 1 & 2 & 3 & 4  \\
\hline
averaged signal $\bar S$ (case A) [$10^1$ cts/epoch] & 40.9 & 40.3 & 36.9 & 40.9 \\
averaged signal $\bar S$ (case B) [$10^1$ cts/epoch] & 10.2 & 10.1 & 9.2 & 10.2  \\
exozodi $N_z$ [$10^3$ cts/epoch] & 12.4 & 9.2 & 7.0 & 4.0  \\
dark noise $N_d$ [$10^3$ cts/epoch] & \multicolumn{4}{c}{1.2}\\
read noise $N_d$ [$10^3$ cts/epoch] & \multicolumn{4}{c}{1.3} \\
averaged $S/N$  (case A) [1/epoch]  $\,^\ast$  & 3.3 & 3.6 & 3.7 & 4.9   \\
averaged $S/N$ (case B)  [1/epoch]  $\,^\ast$ & 0.83 & 0.93 & 0.94 & 1.3   \\
\hline
{\hbox to 0pt{\parbox{200mm}{\footnotesize
$\ast$  The averaged $S/N$ are defined by $\bar{S}/\sqrt{\bar{S}+N_z+N_d+N_r}$ 
   }\hss}}
\end{tabular}
\end{center}
\end{table*}

\begin{figure*}
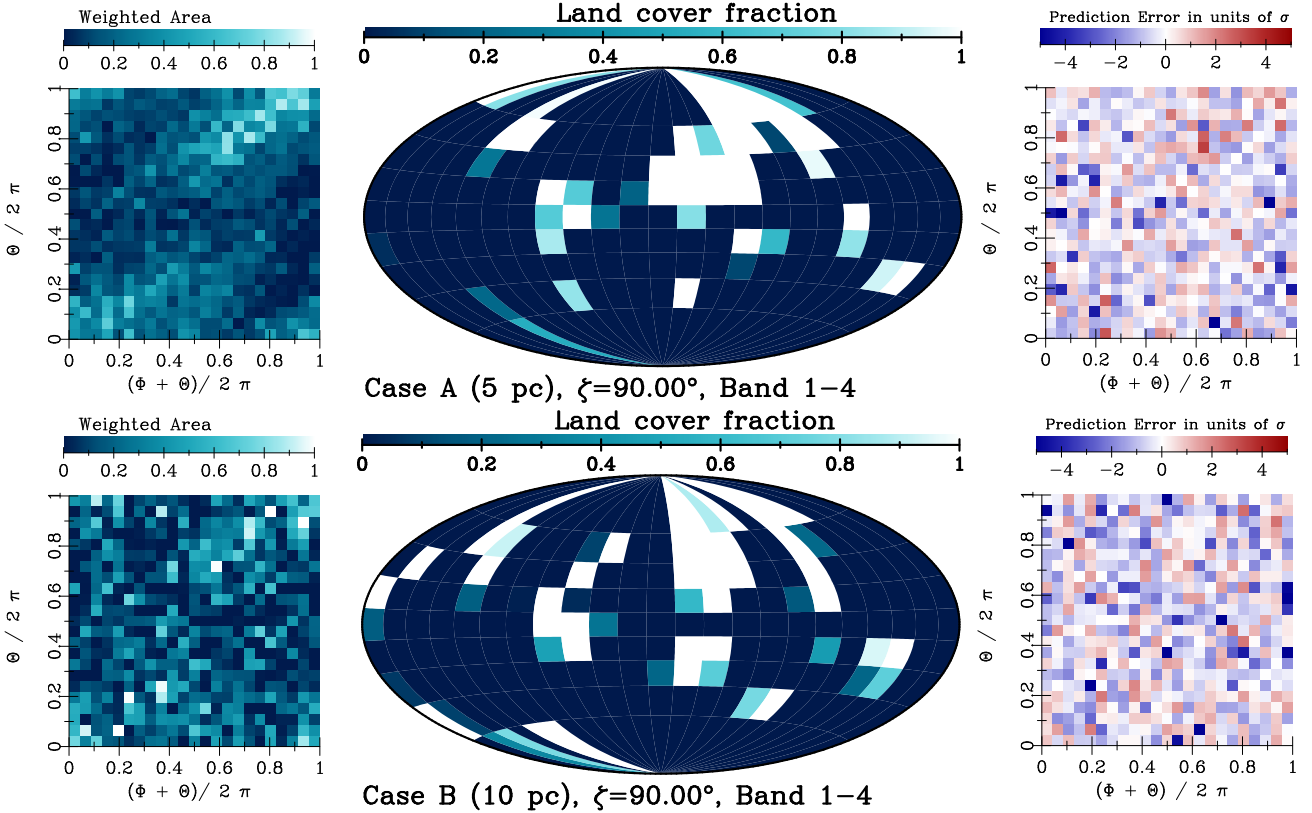

  \begin{minipage}{0.25\linewidth}
    \begin{center}
      \includegraphics[width=\linewidth]{amap_self1-4-O3_5pc90S.ps}
    \end{center}
  \end{minipage}
  \begin{minipage}{0.45\linewidth}
    \begin{center}
      \includegraphics[width=\linewidth]{recover_varextreme_fujii1-90.00b1-4-O3_5pc90S.ps}
    \end{center}
  \end{minipage}
  \begin{minipage}{0.25\linewidth}
    \begin{center}
      \includegraphics[width=\linewidth]{amap_preerr_fujii1-90.00b1-4-O3_5pc90S.ps}
    \end{center}
  \end{minipage}
  \begin{minipage}{0.02\linewidth}
    \begin{center}
      \vspace{0.5cm}
    \end{center}
  \end{minipage}
  \begin{minipage}{0.25\linewidth}
    \begin{center}
      \includegraphics[width=\linewidth]{amap_self1-4-O3_10pc90S.ps}
    \end{center}
  \end{minipage}
  \begin{minipage}{0.45\linewidth}
    \begin{center}
      \includegraphics[width=\linewidth]{recover_varextreme_fujii1-90.00b1-4-O3_10pc90S.ps}
    \end{center}
  \end{minipage}
  \begin{minipage}{0.25\linewidth}
    \begin{center}
      \includegraphics[width=\linewidth]{amap_preerr_fujii1-90.00b1-4-O3_10pc90S.ps}
    \end{center}
  \end{minipage}
  \begin{minipage}{0.02\linewidth}
    \begin{center}
      \vspace{0.5cm}
    \end{center}
  \end{minipage}
 \caption{Simulated weighted area from mock photometric observation (left), its recovered map (middle), and prediction error (right). Top and bottom panels indicates the case A and B,respectively. We note that the snow component such as the South Pole in the input data is replaced to the ocean. \label{fig:fujii-map}}
\end{figure*}

The obtained mock data $I(\lambda_b)$ at each epoch $(\Theta(t_i), \Phi(t_i))$ is decomposed into the weighted area of four surface types, ( ocean, vegetation, soil, and snow ) by solving equation (\ref{eq:invs}) following F10. 
The weighted area of land is computed by summing up that of soil, vegetation, and snow, that is, $A(\Theta ,\Phi ) = A_{\mathrm{soil}} (\Theta , \Phi) +  A_{\mathrm{vegetation}} (\Theta , \Phi ) +  A_{\mathrm{snow}} (\Theta , \Phi)$. 
While F10 used nonlinear fitting in order to constrain $A_k$ to be positive, we use the BVLS linear solver instead without upper constraints. 
The resultant $A(\Theta , \Phi)$ as a function of $\Theta $ and $\Phi $ is displayed in the left panels of Figure  \ref{fig:fujii-map}. 

We now apply the reconstruction method described in \S 2 (see eqs [\ref{eq:defagm}][\ref{eq:least_square}][\ref{eq:cond}]) to the set of obtained $A(\Theta , \Phi)$. For the errors $\sigma _i$ in equation (\ref{eq:defagm}), 
we first compute the variance of the weighted area by 100 times resampling of photon counts according to Poisson statistics, and we call such variance as $\sigma_{i,\mathrm{res}}^2$. 
The $\sigma_{i,\mathrm{res}}$, however,  turned out not to relevant for $\sigma _i$ in equation (\ref{eq:defagm}) because it often results in zero or very small errors when $A_i=0$ due to the boundary condition ($A_i \ge 0$). 
Therefore, we substitute an average value of errors $\sigma =\sum_{i=1}^{\Ndata} \sigma_{i, \mathrm{res}}/\Ndata $ to equation (\ref{eq:defagm}) instead. 
For the case that all estimated errors of $A_k$ are positive, the results with $\sigma_{i}=\sigma_{i,\mathrm{res}}$ do not change significantly.
In the fitting, we assume that the local cover fraction of each surface type remains unchanged and thus ignore the seasonal variations of vegetation, soil, and snow. The center and right rows of Figure \ref{fig:fujii-map} display 
recovered maps, and prediction errors for $\zetain=90^\circ$ of the case A and  B. The major features of the surface can be seen in the resultant maps. The averaged prediction errors $ |A_\mathrm{data} (\Phi(t_i), \Theta(t_i)) - G \mest|/\Ndata$ are 0.53 $\langle A_\mathrm{data} \rangle$ (case A) and 0.85 $\langle A_\mathrm{data} \rangle$ (case B) , respectively. 

\begin{figure*}[!tbh]
  \begin{minipage}{0.5\hsize}
    \begin{center}
      \includegraphics[width=\linewidth]{recover_extreme_fujii-90.00b1-4-O3_5pc90S.ps}
    \end{center}
  \end{minipage}
  \begin{minipage}{0.5\hsize}
    \begin{center}
      \includegraphics[width=\linewidth]{recover_extreme_fujii-90.00b1-4-O3_10pc90S.ps}
    \end{center}
  \end{minipage}
  \begin{minipage}{0.5\hsize}
    \begin{center}
      \includegraphics[width=\linewidth]{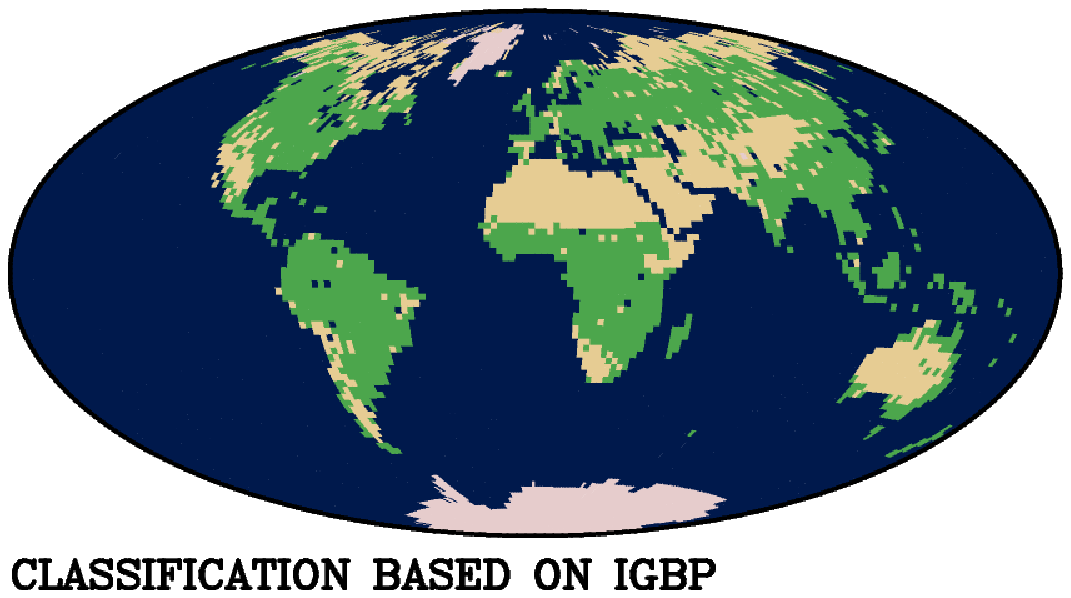}
    \end{center}
  \end{minipage}
\caption{Recovered color maps of soil and vegetation from the mock observations. Soil and vegetation are blended by yellow and green on the land recovered maps shown in the middle panels of Figure \ref{fig:fujii-map}. Top left and right panels indicate the case A and case B, respectively. Snow is replaced to ocean in this simulation. Bottom panel is a classification map of ocean (blue), snow (pink), vegetation (green), and soil (yellow) based on the IGBP classification. Our classification of soil and vegetation is listed in Table 2 of F10. \label{fig:colore}}
\end{figure*}

The land and ocean distributions have been considered so far. Next we create color composition maps of vegetation and soil on the land and ocean recovered map (Figure \ref{fig:colore}).  Adopting the same method of the land recovery to the weighted areas of soil ($A_\mathrm{soil} (t_i)$) and vegetation ($A_\mathrm{vegetation} (t_i)$), we derive the soil and vegetation distributions and blend them by yellow (soil) and green (vegetation) over Figure \ref{fig:fujii-map}, in other words, the land (white) and ocean (blue) distribution. Even using the weighted areas of soil and vegetation separately for the recovery, the resulting vegetation and soil maps almost distribute over the land that recovered with the summation of the weighted areas ($A_\mathrm{land} (t_i) = A_\mathrm{soil} (t_i)+ A_\mathrm{vegetation} (t_i) + A_\mathrm{snow} (t_i)$). In this figure, snow is ignored because the contribution of it is significantly small. The concentrations of vegetation at South America and soil at middle Africa seen in the case A indeed correspond to the Amazon forest and the Sahara desert, respectively. 

\begin{figure*}[!tbh]
  \begin{minipage}{0.95\hsize}
    \begin{center}
      \includegraphics[width=150mm]{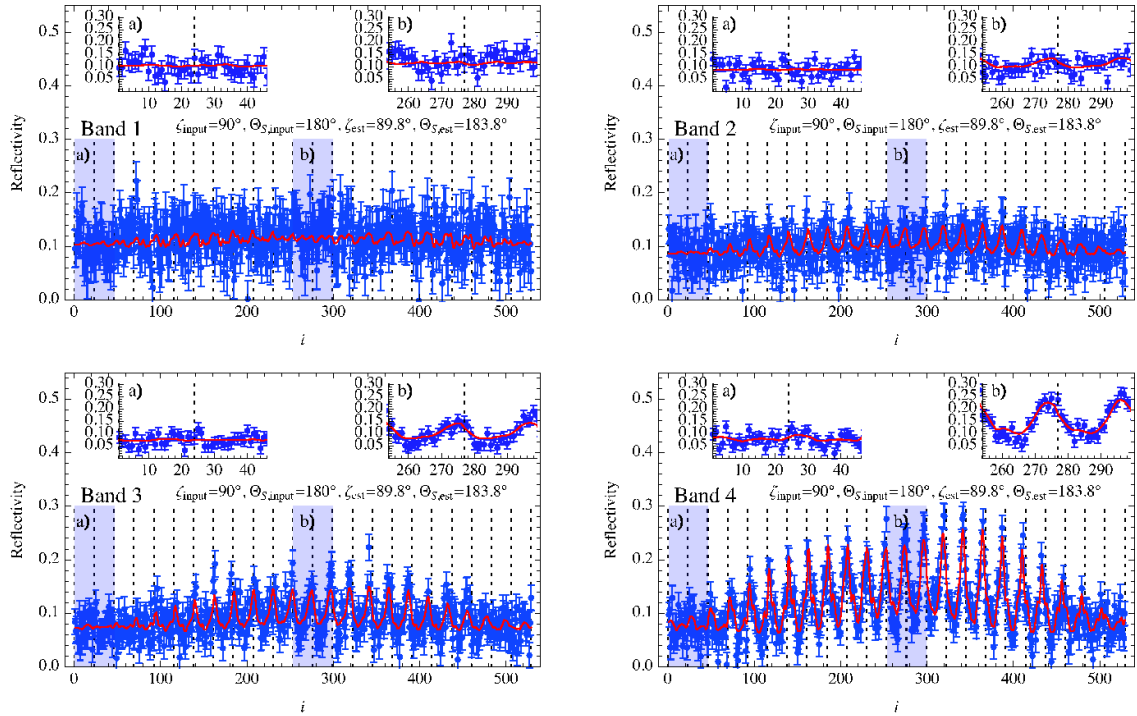}
    \end{center}
  \end{minipage}
  \caption{The mock light curves in units of reflectivity for $\zetain=90^\circ$. Each panel indicates a different band: top left, top right, bottom left, and bottom right panels corresponds to bands 1 ($0.469 \mu$ m), 2 ($0.555 \mu$ m), 3 ($0.645 \mu$ m) and 4 ($0.8585 \mu$ m). Dashed vertical lines divide each epoch $\Phi, \Theta$ stack in 14 days observations. There are 23 data points in one epoch and 23 epochs in whole data set. Therefore, there are $23 \time 23 = 529 $ data points in whole data set ($i=1,2,,,529$).  Wider versions of two shaded regions labeled in a) and b) are inserted in the mini panels. The predicted curves with the best-fit obliquities $\zetaest=89.8^\circ$ and initial longitude $\ThetaSest=183.8$ are drawn by solid lines. The observational noises are computed on the assumption of case A.   \label{fig:lc90}}
\end{figure*}

\begin{figure*}[!tbh]
  \begin{minipage}{0.95\hsize}
    \begin{center}
      \includegraphics[width=150mm]{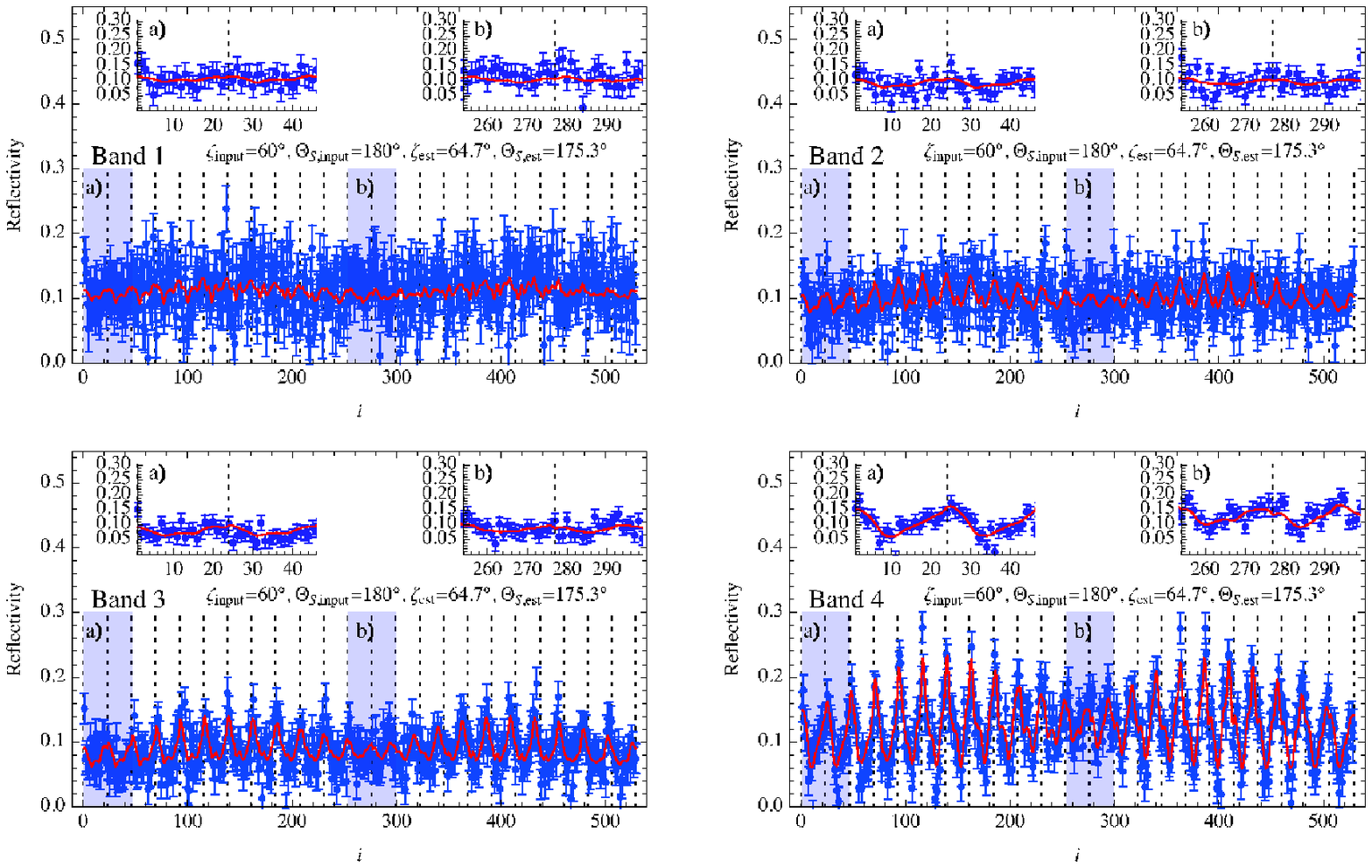}
    \end{center}
  \end{minipage}
  \caption{Same as Figure \ref{fig:lc90} for $\zetain=60^\circ$. \label{fig:lc60}}
\end{figure*}

\begin{figure*}[!tbh]
  \begin{minipage}{0.95\hsize}
    \begin{center}
      \includegraphics[width=150mm]{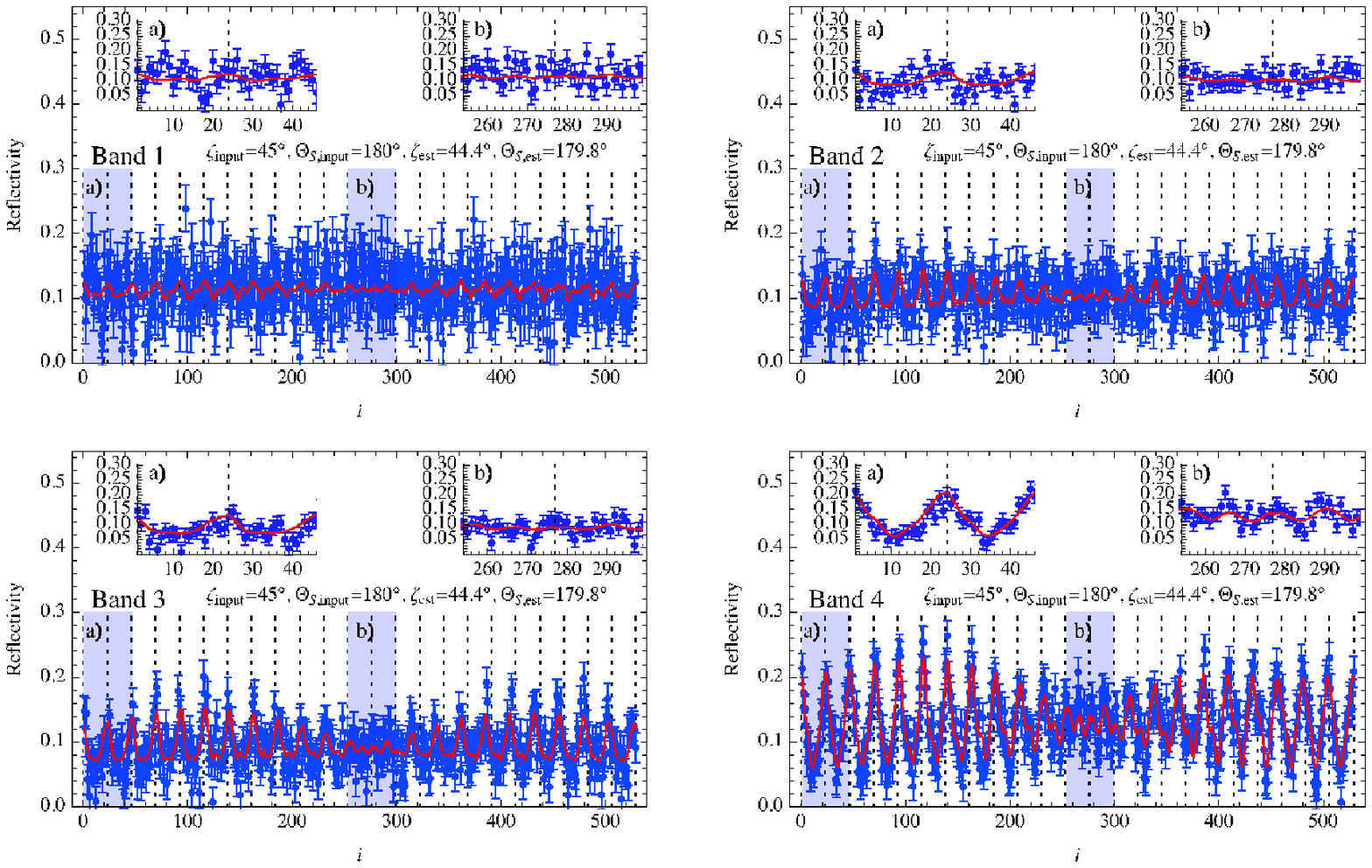}
    \end{center}
  \end{minipage}
  \caption{Same as Figure \ref{fig:lc90} for $\zetain=45^\circ$. \label{fig:lc45}}
\end{figure*}

\begin{figure*}[!tbh]
  \begin{minipage}{0.95\hsize}
    \begin{center}
      \includegraphics[width=150mm]{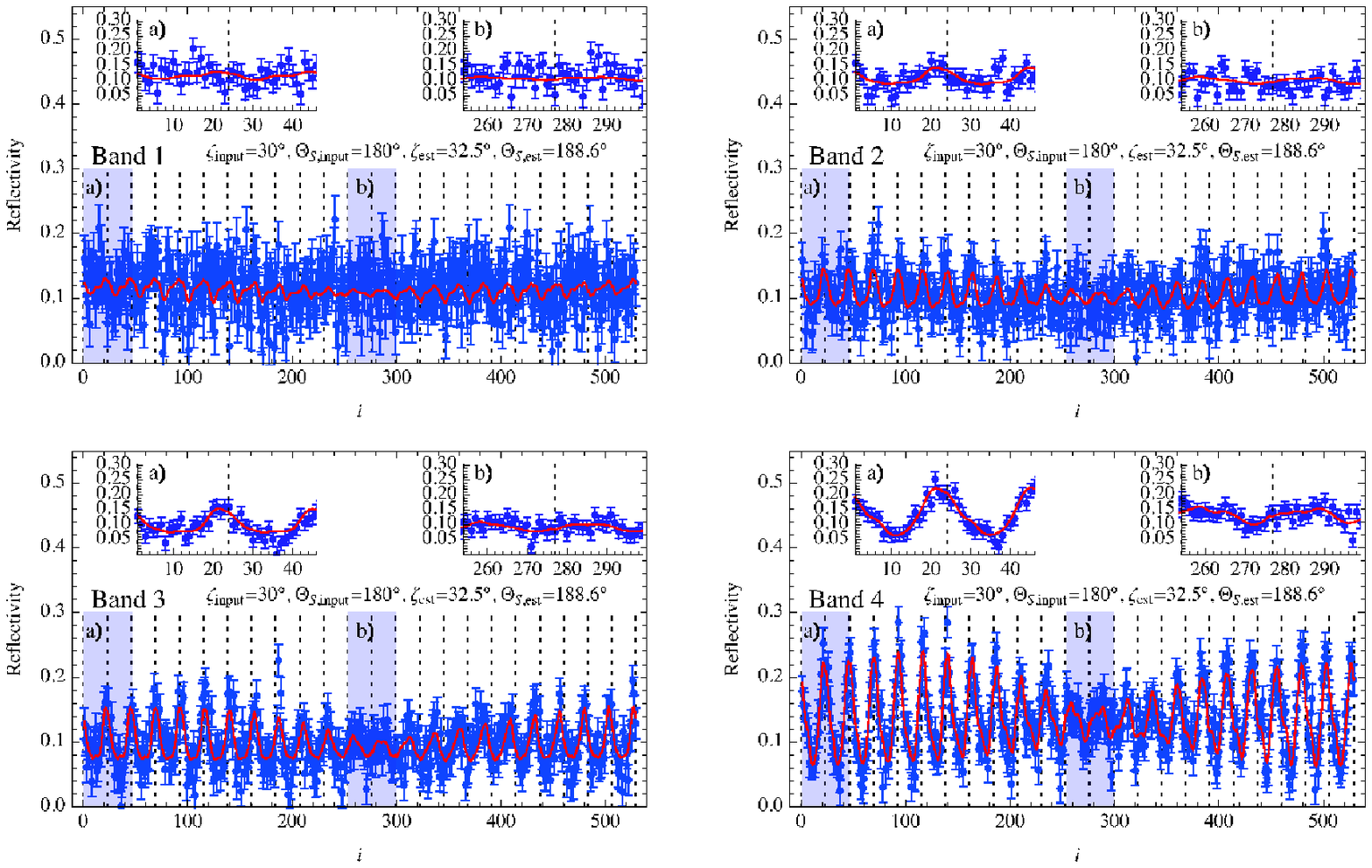}
    \end{center}
  \end{minipage}
  \caption{Same as Figure \ref{fig:lc90} for $\zetain=30^\circ$. \label{fig:lc30}}
\end{figure*}

Now, we try to measure the obliquity for the mock observation. Figures \ref{fig:lc90}, \ref{fig:lc60}, \ref{fig:lc45}, and \ref{fig:lc30} display the light curves of 4 bands with $\ThetaSin=180^\circ$, corresponding to different obliquities, $\zetain=90^\circ, 60^\circ, 45^\circ, $ and $30^\circ$, respectively. The noise assuming in these figures are computed in the ``Case A''. Performing the $\chi^2$ fitting described in \S \ref{cp:zeta}, we estimate the best-fit obliquities and the initial orbital longitude. The predicted curves with the best-fit values are shown by solid lines in Figures \ref{fig:lc90} - \ref{fig:lc30}. The input and the best-fit values and the reduced $\chi^2$ are listed in Table \ref{tab:ob}. We compute the $\chi^2$ for the land distribution (Eq. [\ref{eq:chi2}]). Therefore, the degree of freedom is $23 \times 23 - 23 \times 9 = 322$. It is likely that the inequality constraints and the other systematics such as the Lambert assumption induce the relatively high reduced $\chi^2$ values ($\sim 2$). One can see both diurnal and annual variations of light curves in these figures. The larger variations are seen in band 4 ( 0.8585 $\mu$ m) due to large albedos of the land component (soil and vegetation; see Figure 7 in F10). There is ``pinching'' of the light curve at $\Theta_S = 0^\circ $ ($i \sim 264$) for $\zetain=\pi/4$ because the maximum of the weight function is located at the planet's pole. In this period, the physical position of the maximum stays at the pole for the daily motion. As a result, there are little variations in the period. However the ``pinching'' seen at $\Theta_S \sim 180^\circ (i \sim 0 )$ for $\zetain = 90^\circ$ has a different origin. In this period, the weight function moves around the Southern Hemisphere, that is, ocean mainly.

\begin{figure*}[!tbh]
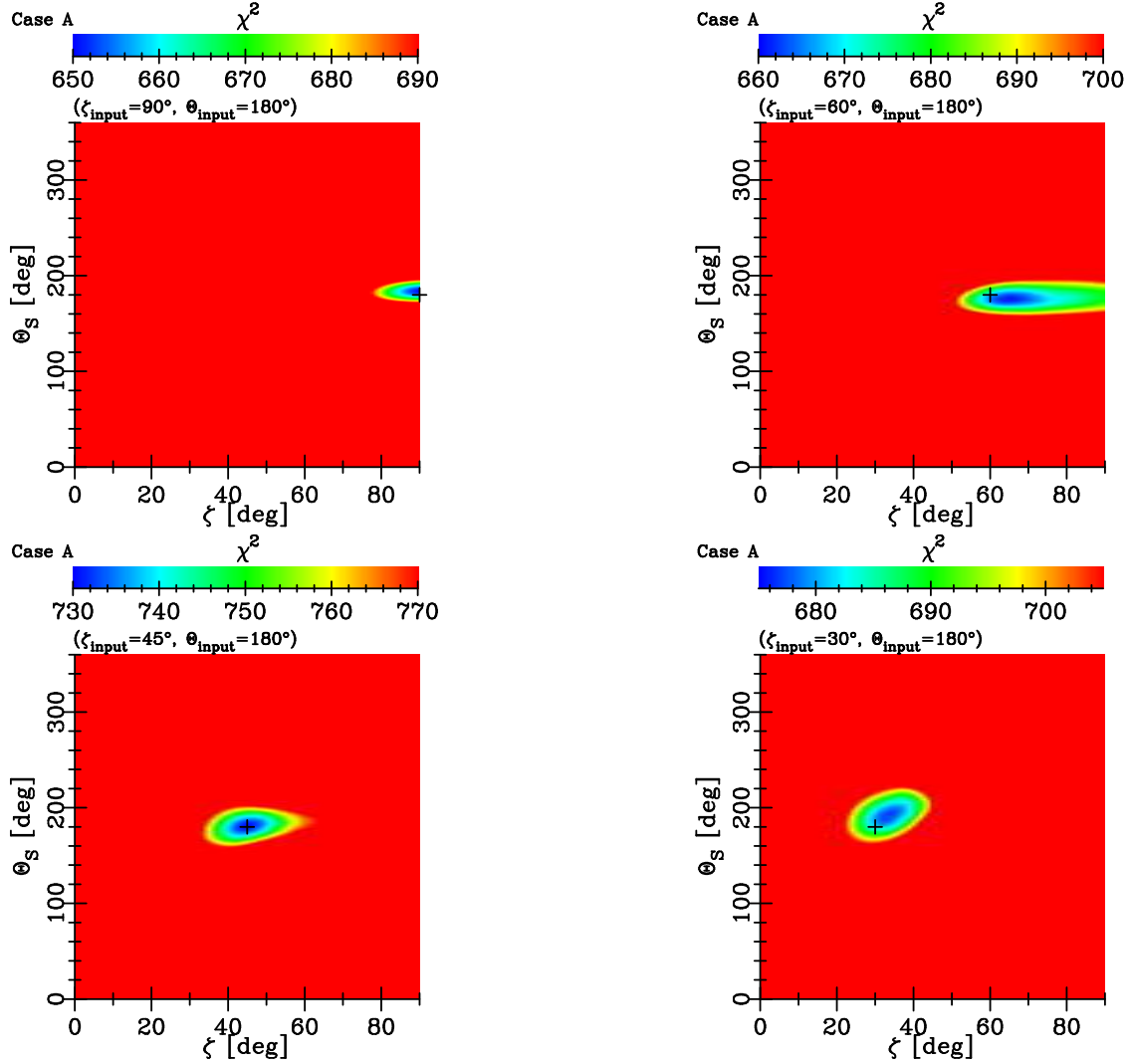

  \begin{minipage}{0.5\hsize}
    \begin{center}
  \includegraphics[width=60mm]{chi-5pc1-490bias.ps}
    \end{center}
  \end{minipage}
  \begin{minipage}{0.5\hsize}
    \begin{center}
  \includegraphics[width=60mm]{chi-5pc1-460bias.ps}
    \end{center}
  \end{minipage}
  \begin{minipage}{0.5\hsize}
    \begin{center}
  \includegraphics[width=60mm]{chi-5pc1-445bias.ps}
    \end{center}
  \end{minipage}
  \begin{minipage}{0.5\hsize}
    \begin{center}
  \includegraphics[width=60mm]{chi-5pc1-430bias.ps}
    \end{center}
  \end{minipage}
  \caption{The $\zeta$ and $\Theta_S$ dependence of $\chi^2(\zeta,\Theta_S)$ of the mock observational data set (case A) for $ \ThetaSin = 180^\circ$ and the different input obliquities, $\zetain = 90^\circ$ (top left)   , $\zetain = 60^\circ$ (top right) , $\zetain = 45^\circ$ (bottom left)  , $\zetain = 30^\circ$ (bottom right). The input values are indicated by cross. The estimated obliquity and initial orbital longitude are listed in Table \ref{tab:ob}. \label{fig:mock-obl}}  
\end{figure*}

\begin{figure*}[!tbh]
\begin{minipage}{0.95\hsize}
    \begin{center}
      \includegraphics[width=150mm]{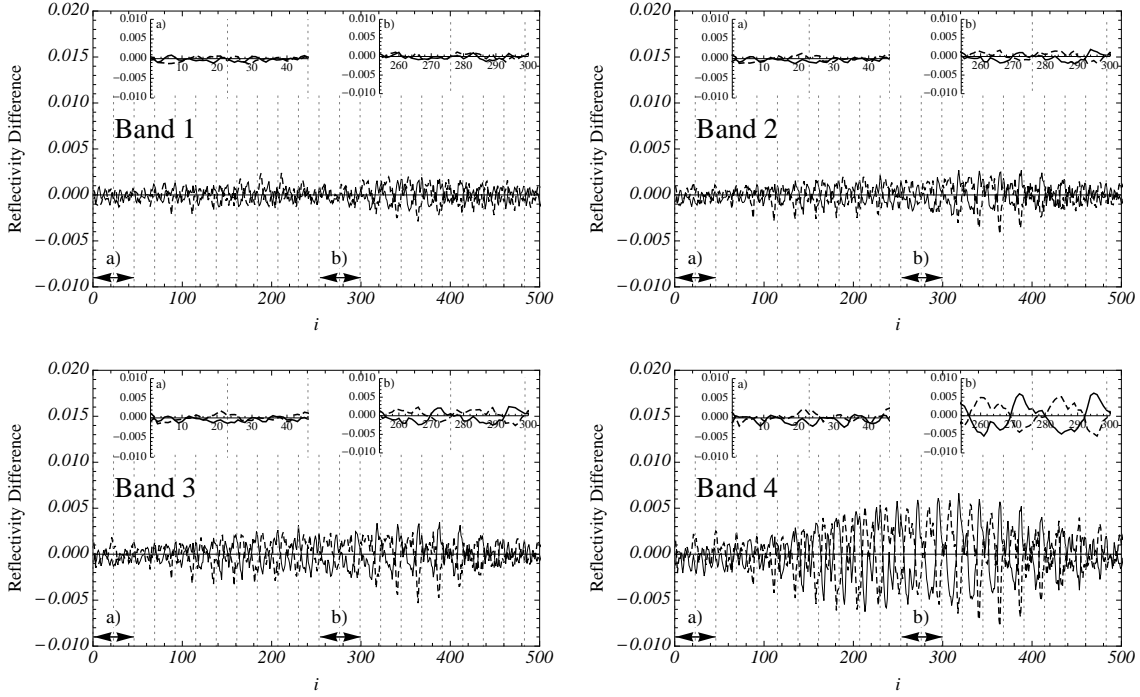}
    \end{center}
  \end{minipage}
  \caption{The difference between the light curve with the best-fit values ($\zetaest=44.4^\circ$, $\ThetaSest=179.8^\circ$), with a $ 3.3^\circ$ shifted obliquity ($\zeta=47.7^\circ, \Theta_S=179.8 ^\circ $; solid curves) and with a $ - 3.3^\circ$ shifted obliquity ($\zeta=41.1^\circ, \Theta_S=179.8^\circ $; dashed curves)    \label{fig:lcd}}
\end{figure*}

The $\chi^2$ maps for the four input obliquities are displayed in Figure \ref{fig:mock-obl} and the best-fit obliquities and initial longitudes are listed in Table \ref{tab:ob}. For the case A, the $\chi^2$ fitting provides the estimated values close to the input ones within $10^\circ$. Figure \ref{fig:lcd} shows the difference of the predicted light curves with the best-fit obliquity $\zetaest=44.4$ and that shifted $\pm 3.3^\circ (\sim 1 \sigma$ level estimated by the bootstrap resampling), $\zeta = 47.7^\circ $ and $41.1^\circ$ . 

\begin{figure*}[!tbh]
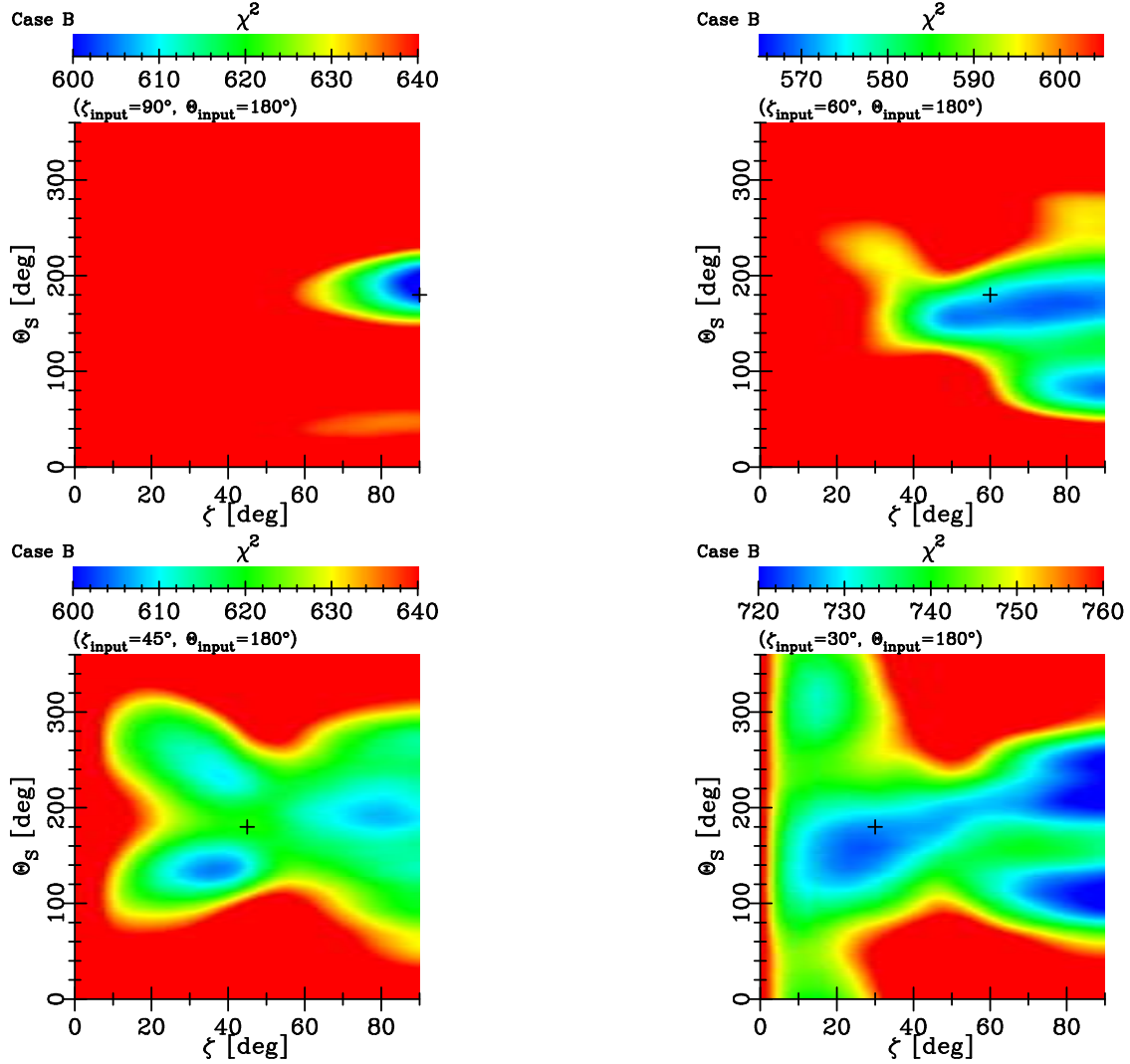

  \begin{minipage}{0.5\hsize}
    \begin{center}
  \includegraphics[width=60mm]{chi-10pc1-490bias.ps}
    \end{center}
  \end{minipage}
  \begin{minipage}{0.5\hsize}
    \begin{center}
  \includegraphics[width=60mm]{chi-10pc1-460bias.ps}
    \end{center}
  \end{minipage}
  \begin{minipage}{0.5\hsize}
    \begin{center}
  \includegraphics[width=60mm]{chi-10pc1-445bias.ps}
    \end{center}
  \end{minipage}
  \begin{minipage}{0.5\hsize}
    \begin{center}
  \includegraphics[width=60mm]{chi-10pc1-430bias.ps}
    \end{center}
  \end{minipage}
  \caption{Same as Figure \ref{fig:mock-obl} for the case B. \label{fig:mock-obl10}}
\end{figure*}

The results for the case B are also listed in Table \ref{tab:ob}. Figure \ref{fig:mock-obl10} displays the $\chi^2$ map for the case B.  The estimated values for $\zetain=30^\circ$ and $60^\circ$ tend to deviate from the input one. As shown in the bottom right panel, the result for $\zetain=30^\circ$ are significantly biased. It is likely that the systematics affect the obliquity estimates for this noise level. We also perform the other method called the extended information criterion in Appendix D. Although the tendency that the estimated value is biased to $\zeta=90^\circ$ vanishes, the best-fit value does not improve significantly.

In short summary, the planetary obliquity can be estimated well for the case A, while, for the case B that has larger noises, it is marginal to derive reliable values of the estimated obliquity by our method.

\begin{table*}[!tbh]
\begin{center}
\caption{Estimated obliquity and the initial orbital longitude by $\chi^2$ minimization and the EIC. \label{tab:ob}}
  \begin{tabular}{lccccccccc}
   \hline\hline 
  &  & & the $\chi^2$ fit & the $\chi^2$ fit &  & & EIC & EIC \\ 
Case   & $\zetain$ & $\ThetaSin$ & $\zetaest$ & $\ThetaSest$ & $\chi^2$ & $\chi^2$/dof & $\zetaest$ & $\ThetaSest$ \\ 
   \hline
 A & $90^\circ$ & $180^\circ$ & $89.8^\circ \pm 1.2^\circ $& $183.8 \pm 2.2^\circ $ & 650.3 & 2.02 & $89.0^\circ$ & $184.5^\circ$\\
 A & $60^\circ$ & $180^\circ$ &$64.7^\circ \pm 4.8^\circ $& $175.3 \pm 3.5^\circ $ & 661.2 & 2.05 & $65.3^\circ$ & $176.5^\circ$ \\
 A & $45^\circ$ & $180^\circ$ &$44.4^\circ \pm 3.3^\circ $& $179.8 \pm 4.1^\circ $ & 731.1 & 2.27 & $44.5^\circ$ & $177.5^\circ$\\
 A & $30^\circ$ & $180^\circ$ &$32.5^\circ \pm 3.0^\circ $& $188.6 \pm 5.6^\circ $ & 677.5 & 2.10 & $30.7^\circ$ & $190.5^\circ$\\
\hline
 B & $90^\circ$ & $180^\circ$ & $89.8^\circ \pm 3.5^\circ$ & $190.8^\circ \pm 17.5^\circ$ & 596.1 & 1.85 & $88.1^\circ$ & $187.5^\circ$\\
 B & $60^\circ$ & $180^\circ$ & $79.7^\circ \pm 8.1^\circ$ & $171.8^\circ \pm 24.8^\circ$ & 568.5 & 1.77 & $48.5^\circ$ & $155.6^\circ$\\
 B & $45^\circ$ & $180^\circ$ & $38.2^\circ \pm 21.6^\circ$ & $231.8^\circ \pm 43.6^\circ$ & 609.5 & 1.89 & $26.7^\circ$ & $132.6^\circ$ \\
 B & $30^\circ$ & $180^\circ$ & $89.2^\circ \pm 17.1^\circ$ & $210.3^\circ \pm 32.7^\circ$ & 714.4 & 2.23 & $7.9^\circ$ & $138.6^\circ$ \\
   \hline
\hline
\end{tabular}
\end{center}
\end{table*}

\section{Summary and Discussion} 
We have developed the reconstruction method of the two-dimensional planetary surface via diurnal and annual variation of the scattered light. Applying the method to the mock photometric data, we have demonstrated that our method works for the mock Earth model, while this model has a lot of simplifying assumptions as follows: 1) cloudless, 2) a face-on circular orbit, 3) known reflectance spectra 4) no atmospheric absorption, 5) known rotation rate 6) static map, and 7) no moon. We also found that the planetary obliquity can be estimated by this method. With our method, future satellite missions such as the occulting ozone observatory \citep{kasdin2010} might provide ``a global map'' of Earth-like exoplanets. While only the terrestrial planets have been considered in this paper, our method might be applicable to any planets with an inhomogeneous surface, including Jupiter-like exoplanets.  

In this paper, we ignored the effect of clouds and expected that clouds affect the estimation as like a statistical noise because of relatively short time variation of clouds. The PCA performed by \cite{2009ApJ...700..915C} is one of promising approaches because they could separate the land and ocean compositions even though they used the EPOXI data that contains the cloud effect. The effect of clouds is discussed elsewhere (Fujii et al. in preparation).¡¡We also assumed a face-on circular orbit in this paper.  This assumption might be too severe for practical applications. We will generalize our method in the next paper.

\acknowledgments 
We are deeply grateful to Yasushi Suto, Atsushi Taruya, and Edwin L. Turner for helpful and insightful discussion. We are thankful to Dmitry Savransky, Jeremy N. Kasdin, and David N. Spergel, who enlightened us about the observing instruments designed for future satellite missions. We also thank the referee, Nicolas Cowan for a lot of constructive comments. HK is supported by a JSPS (Japan Society for Promotion of Science) Grant-in-Aid for science fellows. This work is also supported by Grant-in-Aid for Scientific research from JSPS and from the Japanese Ministry of Education,
Culture, Sports, Science and Technology (Nos. 20$\cdot$10466 and 22$\cdot$5467), and by the
JSPS Core-to-Core Program ``International Research Network for Dark
Energy''.

\begin{appendix}

\section{A. Visible and Illuminated Area \label{ap:iv}}
The visible and illuminated area is expressed as
\begin{eqnarray}
\SVI = \left\{
\begin{array}{c}
 \{(\phi,\theta) | 0 \le \theta \le min[\theta_V (\phi;\Phi;\zeta), \theta_{I,+} (\phi;\Theta; \Phi;\zeta)] \} \nonumber \\
 \mbox{for  $0 < \Theta < \pi/2$ or $3 \pi/2 \le \Theta < \pi$ } , \nonumber \\
 \{(\phi,\theta) | \theta_{I,-} (\phi;\Theta; \Phi;\zeta) \le \theta \le \theta_{V} (\phi;\Phi;\zeta) \} \nonumber \\
\mbox{for  $ \pi/2 \le \Theta < 3 \pi/2$}. \nonumber 
\end{array} \right.\\
\end{eqnarray}
The boundary lines of visible area $ \theta_V (\phi; \Phi;\zeta) $ and illuminated area $ \theta_{I,+} (\phi;\Theta; \Phi;\zeta) $, and $ \theta_{I,-} (\phi;\Theta; \Phi;\zeta) $ are obtained by solving the following equation.
\begin{eqnarray}
&\,& W(\phi,\theta;\Theta;\Phi;\zeta)  \nonumber \\
&=& (\sin{\theta} \cos{(\phi+\Phi)}  \sin{\zeta}  +\cos{\theta} \cos{\zeta} ) \nonumber \\
 &\times& (\cos{\theta} \cos{\Theta} \sin{\zeta} - \sin{\theta} \sin{(\phi+\Phi)} \sin{\Theta}- \sin{\theta} \cos{(\phi+\Phi)} \cos{\Theta} \cos{\zeta}) \nonumber \\
&=& 0.
\end{eqnarray}
The explicit equations of $ \theta_V (\phi; \Phi;\zeta) $, $ \theta_{I,+} (\phi;\Theta; \Phi;\zeta) $, and $ \theta_{I,-} (\phi;\Theta; \Phi;\zeta) $ are expressed as
\begin{eqnarray}
 \theta_V (\phi;\Phi; \zeta) &=& \cos^{-1}{\left( -\frac{\cos{(\phi+\Phi)} \sin{\zeta}}{\sqrt{\cos^2{\zeta} + \cos^2{(\phi+\Phi)} \sin^2{\zeta}}} \right)}, \label{eq:g}
\\
\theta_{I,+} (\phi;\Theta; \Phi;\zeta)  &=& \cos^{-1} (\eta (\phi+\Phi;\Theta;\zeta) ) \\
\theta_{I,-} (\phi;\Theta; \Phi;\zeta)  &=& \cos^{-1} (- \eta (\phi+\Phi;\Theta;\zeta) ),
\end{eqnarray}
where 
\begin{eqnarray}
\eta  (\phi;\Theta;\zeta) &\equiv&\left( \frac{\cos \zeta  \cos \Theta  \cos \phi + \sin \Theta  \sin \phi }{\sqrt{\cos^2 \zeta  \cos^2 \Theta  \cos^2 \phi + 2 \cos \zeta  \sin \Theta  \cos \Theta  \sin \phi  \cos  \phi +\sin^2 \zeta  \cos^2 \Theta +\sin^2 \Theta  \sin^2 \phi }} \right).  \nonumber
\end{eqnarray}

\section{B. Pixelization of Planetary Surface \label{ap:pix}}
We note here that the design matrix alone does not fully specify $\mest$ without any constraints. We perform singular value decomposition (SVD) of the design matrix:
\begin{eqnarray}
G = U \Lambda V^T,
\end{eqnarray}
where $U$ and $V$ are $\Ndata \times \Ndata$ and $\Mmodel \times \Mmodel$ unitary matrices, and $\Lambda$ is a $\Ndata \times \Mmodel$ matrix:
\begin{eqnarray}
\Lambda &=& \left(
\begin{array}{c}
 \Lp  \\
 0 
\end{array}
\right), \\
\Lp &\equiv& diag(\sing_1, .. , \sing_\Mmodel),
\end{eqnarray}
with $\sing_j$ being the singular value of $G$ in descending order. We denote the $j$-th row of $V$ by the vector ${\boldsymbol v_j}$.

In general, there are $\Nnull$ zero components in the singular values, $(\sing_{M-\Nnull+1}=\sing_{M-\Nnull+2}=\sing_{\Mmodel}=0)$. Corresponding ${\boldsymbol v}$ vectors are called null vectors. Linear combination of the null vectors do not affect the data vector,
\begin{eqnarray}
G \sum_{s=1}^{\Nnull} \beta^{\NL,s} {\boldsymbol v}^{\NL,s} &=& {\boldsymbol 0}, \nonumber \\
{\boldsymbol v}^{\NL,s} &\equiv& {\boldsymbol v}_{M-\Nnull+s},
\label{eq:nulleq}
\end{eqnarray}
where $\beta^{\NL,s}$ is an arbitrary coefficient. Equation (\ref{eq:nulleq}) indicates that the predicted model has a freedom to add any linear combination of the null vectors. 

\begin{figure*}[!tbh]
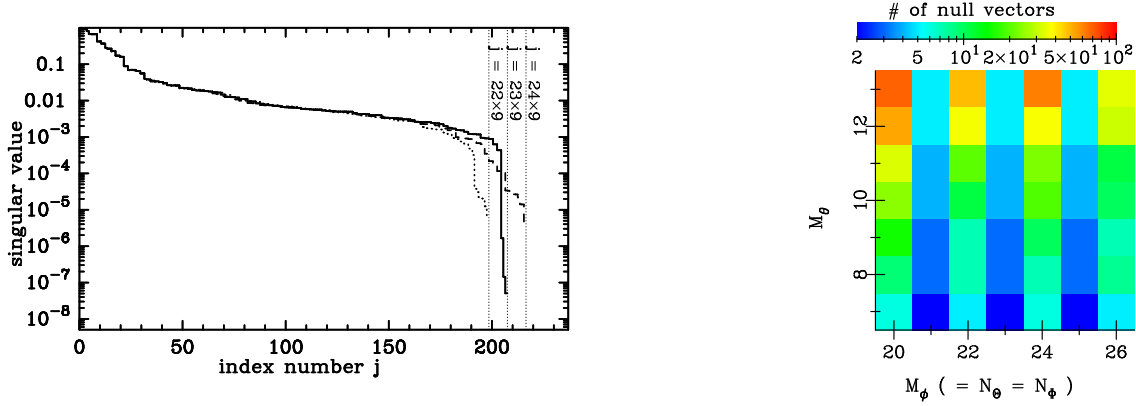

  \begin{minipage}{0.5\hsize}
    \begin{center}
      \includegraphics[width=75mm]{singular_value.ps}
    \end{center}
  \end{minipage}
  \begin{minipage}{0.5\hsize}
    \begin{center}
      \includegraphics[width=45mm]{qnofnullv.ps}
    \end{center}
  \end{minipage}
 \caption{Singular values of the design matrix $G$  in descending order (left panel). Dotted, solid, and dashed curves correspond to $M_\phi=22$, $M_\phi=23$ and $M_\phi=24$, respectively. All curves assume $M_\theta=9$. Right panel shows the number of vectors with singular values below $10^{-4}$ as a function of $M_\theta$ and $M_\phi = N_\Phi = N_\Theta$.  \label{fig:svd} }
\end{figure*}

Solid curve in the left panel of Figure \ref{fig:svd} displays the singular values in descending order (this plot is referred to as {\it spectrum of data kernel}) in the case of $M_{\phi } = N_{\Phi}= N_{\Theta } =23$ and $N_{\theta }=9$. Due to the truncation errors, the singular values obtained by numerical algorithms of the SVD do not vanish exactly. However, only three of them ($j=\Mmodel-2, \Mmodel -1, $ and $\Mmodel$) have significantly smaller values than other diagonal elements. We regard the elements smaller than $10^{-4}$ as zero  and the corresponding row vectors of $V$ as null vectors.  

\begin{figure*}[!tbh]
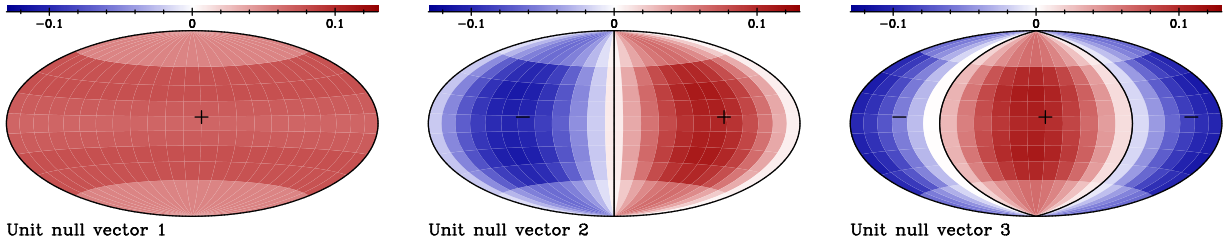

  \begin{minipage}{0.3\hsize}
    \begin{center}
      \includegraphics[width=50mm]{nullvector1.ps}
    \end{center}
  \end{minipage}
  \begin{minipage}{0.3\hsize}
    \begin{center}
      \includegraphics[width=50mm]{nullvector2.ps}
    \end{center}
  \end{minipage}
  \begin{minipage}{0.3\hsize}
    \begin{center}
      \includegraphics[width=50mm]{nullvector3.ps}
    \end{center}
  \end{minipage}
\caption{Three unit null vectors of our fiducial model $M_\phi=N_\Phi=N_\Theta=23$ and $M_\theta=9$. These maps are drawn using Hammer-Aitoff projection. \label{fig:nullvector}}
\end{figure*}

Although the presence of null vectors does not directly lead to the indefiniteness if one adopts the inequality constraints (eq.[\ref{eq:cond}]), smaller number of null vectors makes easier to interpret an uncertainty of the recovered map. Therefore it is useful to know how the number of null vectors depends on pixelization. Then one can decrease the number of the null vectors by an appropriate choice of pixelization. For simplicity, we fix $M_\phi = N_\Phi = N_\Theta$ and estimate the number of the null vectors by changing $M_\theta$ and $M_\phi$. The left panel in Figure \ref{fig:svd} shows the spectra of the singular value in descending order for three cases; $M_\phi=22$ (dotted), $M_\phi=23$ (solid) and $M_\phi=24$ (dashed) with $M_\theta=9$. The case of $M_\phi=23$ has a steeper spectrum than the others. It means that the choice of $M_\phi=23$ makes easier to identify null vectors. The right panel of Figure \ref{fig:svd} indicates the number of null vectors for different sets of $M_\phi$ and $M_\theta$, where we define vectors with singular value below $10^{-4}$ as null vectors. There is a general tendency that the odd number bin of $M_\phi$ has lesser $\Nnull$. Although the selection of smaller values of $M_\theta$ or $M_\phi$ generally leads to smaller number of the null vectors, it degrades the resolution of the map. As a compromise, we decided to adopt $M_\phi=23$ and $M_\theta=9$ for our fiducial values of the model in this paper. Figure \ref{fig:nullvector} indicates the (unit) null vectors for $M_\phi=23$ and $M_\theta=9$, ${\boldsymbol v}^{\NL,1} = {\boldsymbol v}_{\Mmodel-2}, {\boldsymbol v}^{\NL,2} = {\boldsymbol v}_{{\Mmodel-1}}$, and ${\boldsymbol v}^{\NL,3} = {\boldsymbol v}_{\Mmodel}$ where ${\boldsymbol v}_j$ is the $j$-th row of $V$. These three null vectors for $M_\phi=23$ and $M_\theta=9$ affect the large-scale structure of the map but do not change the small-scale distribution like the continents of the present Earth.

\section{C. Uniqueness of the Recovered Map \label{ap:cons}}
 In this appendix, we examine whether there is the freedom for adding any components of null vectors to the recovered map derived by the BVLS. Because of the boundary condition for $\mest$ (eq. [\ref{eq:cond}]), the estimated values of $m_{{\rm est}\,j}$ are classified into 3 types ---(a) $m_{\mathrm{est}\,j} =1$ (on the upper boundary), (b) $m_{{\rm est}\,j} =0$ (on the lower boundary), and (c) 0$<m_{{\rm est}\,j}<1$ (in between).
We focus on types (a) and (b) to consider the uniqueness of our recovered map. 
We denote the index $j$ of type (a) by $j-$ and that of type (b) by $j+$. 
In order not to violate the boundary condition, the linear combination of null vectors should satisfy the following constraints:
\begin{eqnarray}
\sum_{k=1}^{3}\; \beta^{\NL,k} v_{j-}^{\NL,k} &\le& 0 \mbox{\;\;\;for any $j-$},\\
\sum_{k=1}^{3}\; \beta^{\NL,k} v_{j+}^{\NL,k} &\ge& 0 \mbox{\;\;\;for any $j+$}.
\end{eqnarray}
These inequalities can be solved in terms of $\beta^{\NL,1}$ using the fact that $v_j$ is positive (see Fig. [\ref{fig:nullvector}]):
\begin{eqnarray}
\beta^{\NL,1} &\le&  -\frac{v_{j-}^{\NL,2}}{v_{j-}^{\NL,1}} \beta^{\NL,2} -\frac{v_{j-}^{\NL,3}}{v_{j-}^{\NL,1}} \beta^{\NL,3}, \label{eq:beta1}\\
\beta^{\NL,1} &\ge&  -\frac{v_{j+}^{\NL,2}}{v_{j+}^{\NL,1}} \beta^{\NL,2} -\frac{v_{j+}^{\NL,3}}{v_{j+}^{\NL,1}} \beta^{\NL,3}. \label{eq:beta2}
\end{eqnarray}
Then, the following inequality is required so that $\beta^{\NL,1}$ exists, 
\begin{equation}
\frac{v_{j-}^{\NL,2}}{v_{j-}^{\NL,1}} \beta^{\NL,2} + \frac{v_{j-}^{\NL,3}}{v_{j-}^{\NL,1}} \beta^{\NL,3} \le \frac{v_{j+}^{\NL,2}}{v_{j+}^{\NL,1}} \beta^{\NL,2} +\frac{v_{j+}^{\NL,3}}{v_{j+}^{\NL,1}} \beta^{\NL,3}.
\end{equation}
This equation can be reduced to the following expression:
\begin{equation}
{\bf p}_{(j+, j-)} \cdot {\bf b} \ge 0
\label{eq:alphabeta}
\end{equation}
where ${\bf p}_{(j+, j-)}$ and ${\bf b }$ are 2-dimensional vectors:
\begin{eqnarray}
{\bf p}_{(j+, j-)} &\equiv& \left( \frac{v_{j+}^{\NL,2}}{v_{j+}^{\NL,1}} - \frac{v_{j-}^{\NL,2}}{v_{j-}^{\NL,1}}, \frac{v_{j+}^{\NL,3}}{v_{j+}^{\NL,1}} - \frac{v_{j-}^{\NL,3}}{v_{j-}^{\NL,1}} \right)\\
{\bf b } &\equiv& \left( \beta ^{\NL,2}, \beta ^{\NL,3}\right) 
\end{eqnarray}
Now let us denote the arguments of ${\bf p }$ and ${\bf b }$ by $\theta _{\alpha }(j+, j-)$ and $\theta _{\beta }$, respectively, i.e.,
\begin{eqnarray}
{\bf p }_{(j+, j-)} &\equiv& | {\bf p}_{(j+, j-)} | \; ( \cos \theta _{\alpha }(j+, j-), \sin \theta _{\alpha }(j+, j-)), \\
{\bf b } &\equiv& | {\bf b} | \; ( \cos \theta _{\beta }, \sin \theta _{\beta }).
\end{eqnarray}
If $| {\bf b} | \not = 0$, the solution of equation (\ref{eq:alphabeta}) should satisfy
\begin{equation}
\theta _{\alpha }(j+, j-) \le \theta _{\beta } \le \theta _{\alpha }(j+, j-). 
\label{eq:alphabeta2}
\end{equation}
We calculate the above constraints for any combination of $(j+, j-)$, and found that there are no $\theta _{\beta }$ that satisfies equation (\ref{eq:alphabeta2}) for all combinations of $(j+, j-)$. 
Thus, the only solution allowed is $\beta ^{\NL,2} = \beta ^{\NL,3}= 0$.
From equations (\ref{eq:beta1}) and  (\ref{eq:beta2}), it follows that $\beta ^{\NL,1} = 0$. As a result, we confirmed that there is no room for adding extra components of null vectors to the recovered map.  

\section{D. Analysis with the Extended Information Criterion \label{ap:EIC}}

In this appendix, we try to perform the model selection of $\zetaest$ and $\ThetaSest$ using the extended information criterion \citep[EIC; e.g.][]{efron1983,KK1996,ISK1997} based on the Kullback-Leibler divergence. Although the model selection for $\zeta$ and $\Theta_S$ based on the $\chi^2$ minimization is very convenient, the result might be biased for large noises compared with signal.  The EIC is a bootstrap-based extension of Akaike Information Criterion defined as 
\begin{eqnarray}
EIC \equiv - 2 \sum_{i=1}^{\Ndata} \log{ \mathcal{F} ( \A |\mest, \zetaest, \ThetaSest)} + 2 b_B,
\end{eqnarray}
where $ \mathcal{F} ( \A |\mest, \zetaest, \ThetaSest)$ indicates the likelihood function. The best model or parameter set are given by minimizing the EIC. The bootstrap estimate of the bias $b_B$ is given by \citep{ISK1997}:
\begin{eqnarray}
b_B = \frac{1}{N_\mathrm{EIC}} \sum_{k=1}^{N_\mathrm{EIC}}
[\log{ \mathcal{F} ( \A^{\ast,k} |\mest^{\ast,k}, \zetaest^{\ast,k}, \ThetaSest^{\ast,k})} 
- \log{ \mathcal{F} ( \A^{\ast,k} |\mest, \zetaest, \ThetaSest)} \nonumber \\
+ \log{ \mathcal{F} ( \A |\mest, \zetaest, \ThetaSest)} 
- \log{ \mathcal{F} ( \A |\mest^{\ast,k}, \zetaest^{\ast,k}, \ThetaSest^{\ast,k})} ],
\end{eqnarray}
where $ \A^{\ast,k}, \mest^{\ast,k}, \zetaest^{\ast,k}$, and $\ThetaSest^{\ast,k}$ indicate the $k$-th bootstrap resample, $\mest$, $\zetaest$, and $\ThetaSest$, and $N_\mathrm{EIC}$ is the number of trials of the bootstrap.
Assuming a Gaussian distribution for $\A$ with $\sigma=1$, we derive
\begin{eqnarray} 
 EIC &=& \left| \A - G(\zeta,\Theta_S) \mest \right|^2 + \Ndata \log{2 \pi} + b_B \nonumber \\
b_B &=& - \frac{1}{2 N_\mathrm{EIC}} \sum_{k=1}^{N_\mathrm{EIC}} 
[
\left| \A^{\ast,k} - G(\zeta^{\ast,k},\Theta_S^{\ast,k}) \mest^{\ast,k} \right|^2 
- \left| \A^{\ast,k} - G(\zeta,\Theta_S) \mest \right|^2  \nonumber \\
&+& \left| \A - G(\zeta,\Theta_S) \mest \right|^2 
- \left| \A - G(\zeta^{\ast,k},\Theta_S^{\ast,k}) \mest^{\ast,k} \right|^2
].
\label{eq:eic}
\end{eqnarray}

\begin{figure*}[!tbh]
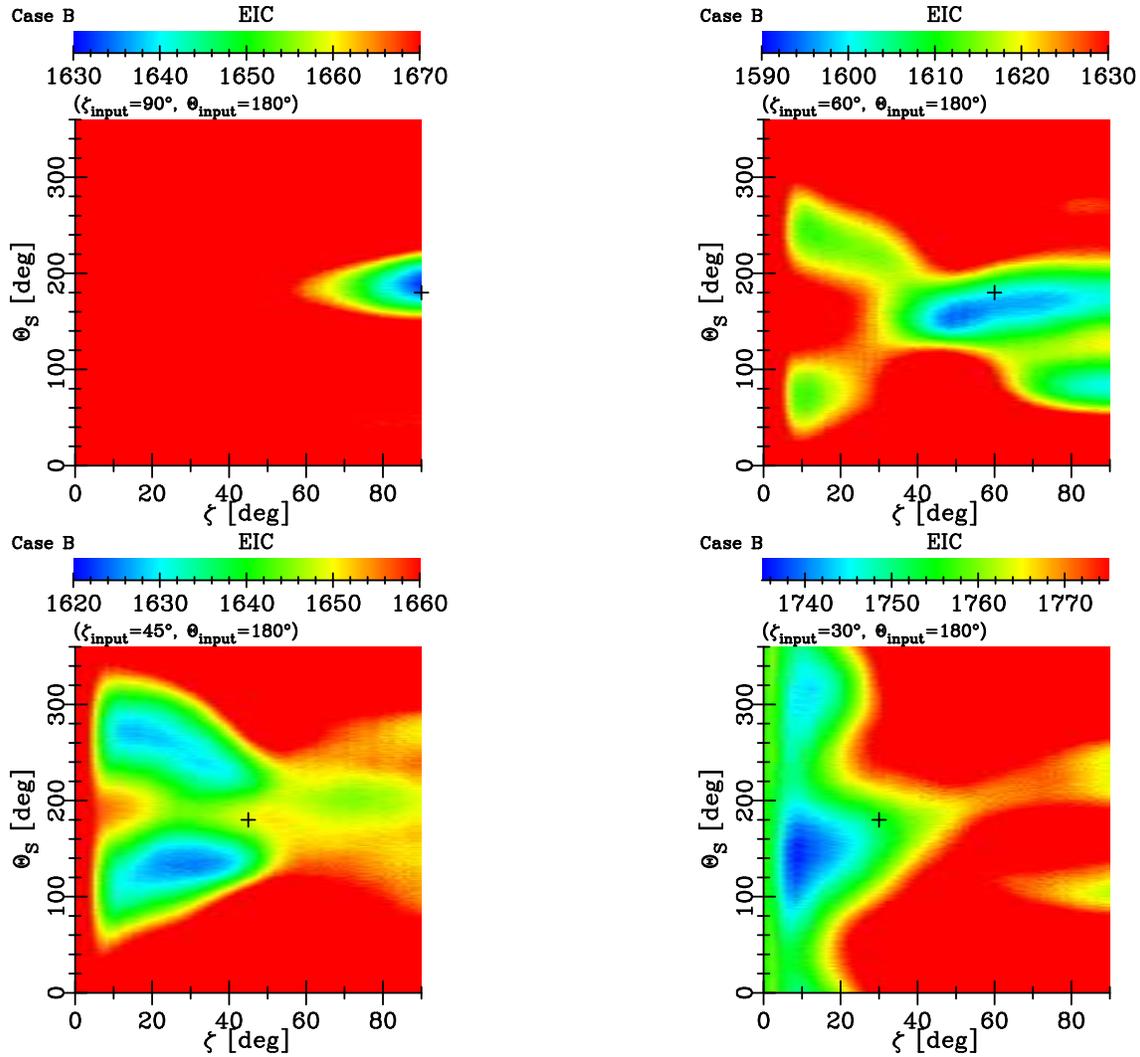

  \begin{minipage}{0.5\hsize}
    \begin{center}
  \includegraphics[width=60mm]{eic-10pc1-490bias.ps}
    \end{center}
  \end{minipage}
  \begin{minipage}{0.5\hsize}
    \begin{center}
  \includegraphics[width=60mm]{eic-10pc1-460bias.ps}
    \end{center}
  \end{minipage}
  \begin{minipage}{0.5\hsize}
    \begin{center}
  \includegraphics[width=60mm]{eic-10pc1-445bias.ps}
    \end{center}
  \end{minipage}
  \begin{minipage}{0.5\hsize}
    \begin{center}
  \includegraphics[width=60mm]{eic-10pc1-430bias.ps}
    \end{center}
  \end{minipage}
  \caption{The $\zeta$ and $\Theta_S$ dependence of the EIC of the mock observational data set (case B) for $ \ThetaSin = 180^\circ$ and the different input obliquities, $\zetain = 90^\circ$ (top left)   , $\zetain = 60^\circ$ (top right) , $\zetain = 45^\circ$ (bottom left)  , $\zetain = 30^\circ$ (bottom right). The input values are indicated by cross. \label{fig:mock-obleic}}  
\end{figure*}

We adopt $N_\mathrm{EIC}=500$ for bootstrap resampling. Figure \ref{fig:mock-obleic} display the dependence of the EIC on $\zetaest$ and $\ThetaSest$.  The minimum EIC estimate listed in Table \ref{tab:ob} is derived by a stepwise search at one degree interval. 

  The minimum EIC estimate listed in Table \ref{tab:ob} is derived by a stepwise search at one degree interval. For the case A, the EIC provides the almost same values as listed in Table \ref{tab:ob}. Figure \ref{fig:mock-obleic} displays the EIC map for the case B. Although the estimated obliquities for $\zetain=30^\circ, 45^\circ$ and $60^\circ$ by the EIC do not agree well with the input value (Table 4), the bias of $\zetain=60^\circ$ and $30^\circ$ seems to be improved compared to that from $\chi^2$ fitting.

\newpage

\begin{thebibliography}{41}
\expandafter\ifx\csname natexlab\endcsname\relax\def\natexlab#1{#1}\fi

\bibitem[{{Agnor} {et~al.}(1999){Agnor}, {Canup}, \&
  {Levison}}]{1999Icar..142..219A}
{Agnor}, C.~B., {Canup}, R.~M., \& {Levison}, H.~F. 1999, Icarus, 142, 219

\bibitem[{{Arnold} {et~al.}(2002){Arnold}, {Gillet}, {Lardi{\`e}re}, {Riaud},
  \& {Schneider}}]{arnold2002}
{Arnold}, L., {Gillet}, S., {Lardi{\`e}re}, O., {Riaud}, P., \& {Schneider}, J.
  2002, \aap, 392, 231

\bibitem[{{Chambers}(2001)}]{2001Icar..152..205C}
{Chambers}, J.~E. 2001, Icarus, 152, 205

\bibitem[{{Charbonneau} {et~al.}(2009){Charbonneau}, {Berta}, {Irwin}, {Burke},
  {Nutzman}, {Buchhave}, {Lovis}, {Bonfils}, {Latham}, {Udry}, {Murray-Clay},
  {Holman}, {Falco}, {Winn}, {Queloz}, {Pepe}, {Mayor}, {Delfosse}, \&
  {Forveille}}]{2009Natur.462..891C}
{Charbonneau}, D., {Berta}, Z.~K., {Irwin}, J., {Burke}, C.~J., {Nutzman}, P.,
  {Buchhave}, L.~A., {Lovis}, C., {Bonfils}, X., {Latham}, D.~W., {Udry}, S.,
  {Murray-Clay}, R.~A., {Holman}, M.~J., {Falco}, E.~E., {Winn}, J.~N.,
  {Queloz}, D., {Pepe}, F., {Mayor}, M., {Delfosse}, X., \& {Forveille}, T.
  2009, \nat, 462, 891

\bibitem[{{Charbonneau} {et~al.}(2002){Charbonneau}, {Brown}, {Noyes}, \&
  {Gilliland}}]{charbonneau2002}
{Charbonneau}, D., {Brown}, T.~M., {Noyes}, R.~W., \& {Gilliland}, R.~L. 2002,
  \apj, 568, 377

\bibitem[{{Colina} {et~al.}(1996){Colina}, {Bohlin}, \&
  {Castelli}}]{colina1996}
{Colina}, L., {Bohlin}, R.~C., \& {Castelli}, F. 1996, {Absolute flux
  calibrated spectrum of Vega} (Instrum. Sci. Rep. OSG-CAL-96-01, Baltimore:
  STSCI)

\bibitem[{{Cowan} \& {Agol}(2008)}]{2008ApJ...678L.129C}
{Cowan}, N.~B., \& {Agol}, E. 2008, \apjl, 678, L129

\bibitem[{{Cowan} {et~al.}(2009){Cowan}, {Agol}, {Meadows}, {Robinson},
  {Livengood}, {Deming}, {Lisse}, {A'Hearn}, {Wellnitz}, {Seager},
  {Charbonneau}, \& {the EPOXI Team}}]{2009ApJ...700..915C}
{Cowan}, N.~B., {Agol}, E., {Meadows}, V.~S., {Robinson}, T., {Livengood},
  T.~A., {Deming}, D., {Lisse}, C.~M., {A'Hearn}, M.~F., {Wellnitz}, D.~D.,
  {Seager}, S., {Charbonneau}, D., \& {the EPOXI Team}. 2009, \apj, 700, 915

\bibitem[{{Efron}(1983)}]{efron1983}
{Efron}, B. 1983, J. Amer. Statist. Assoc, 78, 316

\bibitem[{{Ford} {et~al.}(2001){Ford}, {Seager}, \&
  {Turner}}]{2001Natur.412..885F}
{Ford}, E.~B., {Seager}, S., \& {Turner}, E.~L. 2001, \nat, 412, 885

\bibitem[{{Fujii} {et~al.}(2010){Fujii}, {Kawahara}, {Suto}, {Taruya},
  {Fukuda}, {Nakajima}, \& {Turner}}]{Fujii2010}
{Fujii}, Y., {Kawahara}, H., {Suto}, Y., {Taruya}, A., {Fukuda}, S.,
  {Nakajima}, T., \& {Turner}, E.~L. 2010, \apj, 715, 866

\bibitem[{{Gaidos} \& {Williams}(2004)}]{2004NewA...10...67G}
{Gaidos}, E., \& {Williams}, D.~M. 2004, New Astronomy, 10, 67

\bibitem[{{Ishiguro} {et~al.}(1997){Ishiguro}, {Sakamoto}, \&
  {Kitagawa}}]{ISK1997}
{Ishiguro}, M., {Sakamoto}, Y., \& {Kitagawa}, G. 1997, Ann. Inst. Statist.
  Math., 49, 411

\bibitem[{{Kasdin} {et~al.}(2010){Kasdin}, {Spergel}, {Vanderbei}, {Cady},
  {Savransky}, {Lisman}, {Shaklan}, {Lee}, {Egerman}, {Matthews}, \&
  {Tenerelli}}]{kasdin2010}
{Kasdin}, N.~J., {Spergel}, D.~N., {Vanderbei}, R.~J., {Cady}, E., {Savransky},
  D., {Lisman}, D., {Shaklan}, S., {Lee}, R., {Egerman}, R., {Matthews}, G., \&
  {Tenerelli}, D. 2010, in Bulletin of the American Astronomical Society,
  Vol.~41, Bulletin of the American Astronomical Society, 287--+

\bibitem[{{Knutson} {et~al.}(2007){Knutson}, {Charbonneau}, {Allen}, {Fortney},
  {Agol}, {Cowan}, {Showman}, {Cooper}, \& {Megeath}}]{2007Natur.447..183K}
{Knutson}, H.~A., {Charbonneau}, D., {Allen}, L.~E., {Fortney}, J.~J., {Agol},
  E., {Cowan}, N.~B., {Showman}, A.~P., {Cooper}, C.~S., \& {Megeath}, S.~T.
  2007, \nat, 447, 183

\bibitem[{{Kokubo} \& {Ida}(2007)}]{2007ApJ...671.2082K}
{Kokubo}, E., \& {Ida}, S. 2007, \apj, 671, 2082

\bibitem[{{Konishi} \& {Kitagawa}(1996)}]{KK1996}
{Konishi}, S., \& {Kitagawa}, G. 1996, Biometrika, 83, 875

\bibitem[{{Lawson} \& {Hanson}(1974)}]{1974slsp.book.....L}
{Lawson}, C.~L., \& {Hanson}, R.~J. 1974, {Solving least squares problems}

\bibitem[{{Lawson} \& {Hanson}(1995)}]{1995LH}
---. 1995, {Solving least squares problems (Revised edition)} (Society for
  Industrial and Applied Mathematics)

\bibitem[{{L{\'e}ger} {et~al.}(2009){L{\'e}ger}}]{2009A&A...506..287L}
{L{\'e}ger}, A. et al. 2009, \aap, 506, 287

\bibitem[{{Menke}(1989)}]{1989gdad.book.....M}
{Menke}, W. 1989, {Geophysical data analysis: Discrete inverse theory}, ed.
  {Menke, W.}

\bibitem[{{Monta{\~n}{\'e}s-Rodr{\'{\i}}guez}
  {et~al.}(2006){Monta{\~n}{\'e}s-Rodr{\'{\i}}guez}, {Pall{\'e}}, {Goode}, \&
  {Mart{\'{\i}}n-Torres}}]{2006ApJ...651..544M}
{Monta{\~n}{\'e}s-Rodr{\'{\i}}guez}, P., {Pall{\'e}}, E., {Goode}, P.~R., \&
  {Mart{\'{\i}}n-Torres}, F.~J. 2006, \apj, 651, 544

\bibitem[{{Nakajima} \& {Tanaka}(1983)}]{nakajima1983}
{Nakajima}, T., \& {Tanaka}, M. 1983, Journal of Quantitative Spectroscopy and
  Radiative Transfer, 29, 521

\bibitem[{{Oakley} \& {Cash}(2009)}]{2009ApJ...700.1428O}
{Oakley}, P.~H.~H., \& {Cash}, W. 2009, \apj, 700, 1428

\bibitem[{{Pall{\'e}} {et~al.}(2008){Pall{\'e}}, {Ford}, {Seager},
  {Monta{\~n}{\'e}s-Rodr{\'{\i}}guez}, \& {Vazquez}}]{2008ApJ...676.1319P}
{Pall{\'e}}, E., {Ford}, E.~B., {Seager}, S.,
  {Monta{\~n}{\'e}s-Rodr{\'{\i}}guez}, P., \& {Vazquez}, M. 2008, \apj, 676,
  1319

\bibitem[{{Press} {et~al.}(1992){Press}, {Teukolsky}, {Vetterling}, \&
  {Flannery}}]{1992nrfa.book.....P}
{Press}, W.~H., {Teukolsky}, S.~A., {Vetterling}, W.~T., \& {Flannery}, B.~P.
  1992, {Numerical recipes in FORTRAN. The art of scientific computing}, ed.
  {Press, W.~H., Teukolsky, S.~A., Vetterling, W.~T., \& Flannery, B.~P.}

\bibitem[{{Russell}(1906)}]{1906ApJ....24....1R}
{Russell}, H.~N. 1906, \apj, 24, 1

\bibitem[{{Savransky} {et~al.}(2010){Savransky}, {Kasdin}, \&
  {Cady}}]{2010PASP..122..401S}
{Savransky}, D., {Kasdin}, N.~J., \& {Cady}, E. 2010, \pasp, 122, 401

\bibitem[{{Seager} {et~al.}(2000){Seager}, {Whitney}, \&
  {Sasselov}}]{2000ApJ...540..504S}
{Seager}, S., {Whitney}, B.~A., \& {Sasselov}, D.~D. 2000, \apj, 540, 504

\bibitem[{{Sudarsky} {et~al.}(2005){Sudarsky}, {Burrows}, {Hubeny}, \&
  {Li}}]{2005ApJ...627..520S}
{Sudarsky}, D., {Burrows}, A., {Hubeny}, I., \& {Li}, A. 2005, \apj, 627, 520

\bibitem[{{Swain} {et~al.}(2008){Swain}, {Vasisht}, \& {Tinetti}}]{swain2008}
{Swain}, M.~R., {Vasisht}, G., \& {Tinetti}, G. 2008, \nat, 452, 329

\bibitem[{{Swain} {et~al.}(2009){Swain}, {Vasisht}, {Tinetti}, {Bouwman},
  {Chen}, {Yung}, {Deming}, \& {Deroo}}]{swain2009}
{Swain}, M.~R., {Vasisht}, G., {Tinetti}, G., {Bouwman}, J., {Chen}, P.,
  {Yung}, Y., {Deming}, D., \& {Deroo}, P. 2009, \apjl, 690, L114

\bibitem[{{Tinetti} {et~al.}(2006{\natexlab{a}}){Tinetti}, {Meadows}, {Crisp},
  {Fong}, {Fishbein}, {Turnbull}, \& {Bibring}}]{2006AsBio...6...34T}
{Tinetti}, G., {Meadows}, V.~S., {Crisp}, D., {Fong}, W., {Fishbein}, E.,
  {Turnbull}, M., \& {Bibring}, J. 2006{\natexlab{a}}, Astrobiology, 6, 34

\bibitem[{{Tinetti} {et~al.}(2006{\natexlab{b}}){Tinetti}, {Meadows}, {Crisp},
  {Kiang}, {Kahn}, {Fishbein}, {Velusamy}, \& {Turnbull}}]{2006AsBio...6..881T}
{Tinetti}, G., {Meadows}, V.~S., {Crisp}, D., {Kiang}, N.~Y., {Kahn}, B.~H.,
  {Fishbein}, E., {Velusamy}, T., \& {Turnbull}, M. 2006{\natexlab{b}},
  Astrobiology, 6, 881

\bibitem[{{Tinetti} {et~al.}(2007){Tinetti}, {Vidal-Madjar}, {Liang},
  {Beaulieu}, {Yung}, {Carey}, {Barber}, {Tennyson}, {Ribas}, {Allard},
  {Ballester}, {Sing}, \& {Selsis}}]{tinetti2007}
{Tinetti}, G., {Vidal-Madjar}, A., {Liang}, M., {Beaulieu}, J., {Yung}, Y.,
  {Carey}, S., {Barber}, R.~J., {Tennyson}, J., {Ribas}, I., {Allard}, N.,
  {Ballester}, G.~E., {Sing}, D.~K., \& {Selsis}, F. 2007, \nat, 448, 169

\bibitem[{{Vidal-Madjar} {et~al.}(2004){Vidal-Madjar}, {D{\'e}sert},
  {Lecavelier des Etangs}, {H{\'e}brard}, {Ballester}, {Ehrenreich}, {Ferlet},
  {McConnell}, {Mayor}, \& {Parkinson}}]{vidal2004}
{Vidal-Madjar}, A., {D{\'e}sert}, J., {Lecavelier des Etangs}, A.,
  {H{\'e}brard}, G., {Ballester}, G.~E., {Ehrenreich}, D., {Ferlet}, R.,
  {McConnell}, J.~C., {Mayor}, M., \& {Parkinson}, C.~D. 2004, \apjl, 604, L69

\bibitem[{{Vidal-Madjar} {et~al.}(2003){Vidal-Madjar}, {Lecavelier des Etangs},
  {D{\'e}sert}, {Ballester}, {Ferlet}, {H{\'e}brard}, \& {Mayor}}]{vidal2003}
{Vidal-Madjar}, A., {Lecavelier des Etangs}, A., {D{\'e}sert}, J., {Ballester},
  G.~E., {Ferlet}, R., {H{\'e}brard}, G., \& {Mayor}, M. 2003, \nat, 422, 143

\bibitem[{{Williams} \& {Kasting}(1997)}]{1997Icar..129..254W}
{Williams}, D.~M., \& {Kasting}, J.~F. 1997, Icarus, 129, 254

\bibitem[{{Williams} \& {Pollard}(2003)}]{2003IJAsB...2....1W}
{Williams}, D.~M., \& {Pollard}, D. 2003, International Journal of
  Astrobiology, 2, 1

\bibitem[{{Williams} {et~al.}(2006){Williams}, {Charbonneau}, {Cooper},
  {Showman}, \& {Fortney}}]{2006ApJ...649.1020W}
{Williams}, P.~K.~G., {Charbonneau}, D., {Cooper}, C.~S., {Showman}, A.~P., \&
  {Fortney}, J.~J. 2006, \apj, 649, 1020

\bibitem[{{Woolf} {et~al.}(2002){Woolf}, {Smith}, {Traub}, \&
  {Jucks}}]{woolf2002}
{Woolf}, N.~J., {Smith}, P.~S., {Traub}, W.~A., \& {Jucks}, K.~W. 2002, \apj,
  574, 430

\end{thebibliography}
\end{appendix}

\end{document}